\documentclass[12pt]{article}
\usepackage[a4paper, top=16mm, text={200mm, 248mm}, includehead, includefoot, hmarginratio=1:1, heightrounded]{geometry}
\usepackage{amsmath,amssymb,mathrsfs,amsthm,tikz,shuffle,paralist}
\usepackage{dsfont}
\usepackage{color}
\definecolor{darkred}{RGB}{173,34,48}

\usetikzlibrary{snakes}
\usetikzlibrary{calc}
\usetikzlibrary{decorations}

\usepackage[all]{xypic}
\usepackage{jheppub}
\usepackage{hyperref}
\usepackage{subfigure}
\usepackage{caption}

\newcommand{\dif}{\mathrm{d}} 

\title{On symbology and differential equations of Feynman integrals from Schubert analysis}
\date{\today}

\author[a,b,c]{Song He,}\emailAdd{songhe@itp.ac.cn}
\author[a]{Xuhang Jiang,}\emailAdd{xhjiang@itp.ac.cn}
\author[a,d]{Jiahao Liu,}\emailAdd{liujiahao@itp.ac.cn}
\author[a,d]{Qinglin Yang}\emailAdd{yangqinglin@itp.ac.cn}
\affiliation[a]{CAS Key Laboratory of Theoretical Physics, Institute of Theoretical Physics, Chinese Academy of Sciences, Beijing 100190, China}
\affiliation[b]{School of Fundamental Physics and Mathematical Sciences, Hangzhou Institute for Advanced Study, UCAS \& ICTP-AP, Hangzhou, 310024, China}
\affiliation[c]{Peng Huanwu Center for Fundamental Theory, Hefei, Anhui 230026, P. R. China}
\affiliation[d]{School of Physical Sciences, University of Chinese Academy of Sciences, No.19A Yuquan Road, Beijing 100049, China}

\abstract{We take the first step in generalizing the so-called ``Schubert analysis", originally proposed in twistor space for four-dimensional kinematics, to the study of symbol letters and more detailed information on canonical differential equations for Feynman integral families in general dimensions with general masses. The basic idea is to work in embedding space and compute possible cross-ratios built from (Lorentz products of) maximal cut solutions for all integrals in the family. We demonstrate the power of the method using the most general one-loop integrals, as well as various two-loop planar integral families (such as sunrise, double-triangle and double-box) in general dimensions. Not only can we obtain all symbol letters as cross-ratios from maximal-cut solutions, but we also reproduce entries in the canonical differential equations satisfied by a basis of ${\rm d} \log$ integrals. 
}
\begin{document}

\maketitle

\section{Introduction}
In~\cite{Yang:2022gko}, a new method was proposed for studying the symbology of multi-loop Feynman integrals in $D=4-2 \epsilon$, which was based on the so-called Schubert problems \cite{Hodges:2010kq,ArkaniHamed:2010gh} for geometric configurations (intersections of lines {\it etc.}) in momentum twistor space~\cite{Hodges:2009hk}. The original idea was proposed in the context of loop amplitudes of ${\cal N}=4$ SYM and corresponding dual-conformal-invariant (DCI) integrals in $D=4$~\cite{Drummond:2006rz,Drummond:2010cz,ArkaniHamed:2010gh,Spradlin:2011wp,DelDuca:2011wh,Bourjaily:2013mma,Henn:2018cdp,Herrmann:2019upk, Bourjaily:2018aeq,Bourjaily:2019hmc,He:2020uxy,He:2020lcu} which are most naturally formulated in momentum twistor space: by computing conformal cross-ratios of maximal cut solutions or \emph{leading singularities} (LS)~\cite{Bern:1994zx,Britto:2004nc,Cachazo:2008vp}, which amounts to intersection points of external and internal (loop) lines in twistor space, one can successfully predict ~\cite{Yang:2022gko,He:2022ctv} the alphabet of the {\it symbol}~\cite{Goncharov:2010jf,Duhr:2011zq} of such DCI integrals. These include well known examples for $n=6,7$, but also the very recent results of $9+9$ algebraic letters for $n=8$ and even higher point cases~\cite{Zhang:2019vnm,He:2020lcu,He:2021non}.

More generally speaking, ${\cal N}=4$ SYM has also become an extremely fruitful laboratory for new methods of evaluating general Feynman integrals. Similar to the full amplitudes, individual Feynman integrals exhibit unexpected mathematical structures such as \emph{cluster algebras}~\cite{fomin2002cluster}, {\it e.g.} $A_3$ and $E_6$ for six- and seven-point amplitudes/integrals in SYM~\cite{Golden:2013xva}. These alphabets are the starting point of bootstrapping such amplitudes to impressively high loop orders ({\it c.f.}~\cite{Dixon:2011pw,Dixon:2014xca,Dixon:2014iba,Drummond:2014ffa,Dixon:2015iva,Caron-Huot:2016owq,Dixon:2016nkn,Drummond:2018caf, Caron-Huot:2019vjl, Caron-Huot:2019bsq} and a review~\cite{Caron-Huot:2020bkp}). It is highly non-trivial that cluster algebras and extensions also control the symbol alphabets of individual (all-loop) Feynman integrals beyond the $n=6,7$ cases~\cite{Caron-Huot:2018dsv, Drummond:2017ssj,He:2021esx,He:2021non}. Even more surprisingly, cluster algebraic structures have been identified and explored for the symbology of more general, non-DCI Feynman integrals in $4-2 \epsilon$ dimensions (with massless propagators)~\cite{Chicherin:2020umh}. It is very interesting that all these cluster-algebraic structures and beyond can be nicely accounted for by the Schubert analysis formulated in momentum twistor space: by introducing the line $I_{\infty}$ which breaks conformal invariance, essentially the same Schubert analysis \cite{He:2022tph} produces symbol alphabets for one- and higher-loop Feynman integrals in $4-2\epsilon$ dimensions with various kinematics such as one-mass five-point case~\cite{Abreu:2020jxa}. In fact, in addition to numerous examples for cases when Feynman integrals evaluate to MPL functions, there is also evidence that Schubert analysis can be extended to the ``symbol letters" for elliptic MPL case, such as double-box integrals in $D=4$~\cite{Morales:2022csr} 
 (see~\cite{He:2023qld} for closely related works).

Despite the success this method has been restricted to planar Feynman integrals near (or in) four dimensions with no internal mass, where momentum twistors are extremely powerful. In particular, on-shell (cut) conditions correspond to intersecting lines (which correspond to dual points in $d=4$) and symbol letters are then produced by {\it cross-ratios} of these intersection points. It is natural to wonder if one could generalize it to Feynman integrals in general dimensions, and/or with general internal masses. At first glance this seems rather difficult: even for planar integrals near other integer dimensions, we do not have such momentum twistor variables (except for $d=3$~\cite{Elvang:2014fja}), and the presence of internal mass makes it unclear how to define geometric configurations for maximal cuts. 

In this paper, we take the first step in generalizing this method to Feynman integrals near $d$ dimensions with possible internal masses in {\it embedding space}, and we focus on planar integrals for now (we will comment on possible extensions in the end). The idea of such generalized Schubert analysis is very similar to the original one in $d=4$: even without any direct geometric interpretations, we can first solve maximal cuts (possibly with internal masses) which are expressed in terms of $(d{+}2)$-dimensional vectors, and then consider cross-ratios of Lorentz products of such solutions. Indeed we will see that for most general family of one-loop integrals and various highly non-trivial two-loop integral families, this Schubert analysis in embedding space works very nicely.

In the rest of this section, we will sketch the basic idea and leave the details to the remaining sections. Before that, let us first briefly review the embedding space formalism for $d$ dimensions. For planar integrals with $ n $ external momenta, we may introduce dual points $x_i^\mu$ such that $p_i=x_i-x_{i+1}$ for $ i=1,\cdots,n \, (x_{n+1}\equiv x_1)$  to trivialize momentum conservation. We further embed the dual spacetime $\mathbb{R}^{d-1,1}$ into $\mathbb{R}^{d,2}$ where the dual conformal group acts linearly:
\begin{equation}\label{embbeddef}
    x_i^\mu\leadsto X_i^M=(x_i^\mu;-x_i^2+m_i^2,1)
\end{equation}
We have written the embedding space vectors in light-cone coordinates, so the dot product of two vectors reads
\begin{align}\label{eq:dotproduct}
   (X_i,X_j):&=\eta_{MN}X_i^MX_j^N=-2X_i^\mu X_{j,\mu}-X_i^+X_j^--X_i^-X_j^+,\\
   &=(x_i-x_j)^2-m_i^2-m_j^2. 
\end{align}
For loop integrals, we assign a dual point $y_k^\mu$ to each loop, such that the propagator between $y_k$ and $x_i$ is $[(y_k-x_i)^2-m_i^2+i0]^{-1}$. Loop momenta correspond to extra vectors in embedding space:
\begin{equation}
y_k^\mu\leadsto Y_k^M=\alpha(y_k^\mu;-y_k^2,1)
\end{equation} 
where parameter $\alpha\neq0$ indicates the projective freedom. It is useful to introduce the ``infinity''  $I_\infty=(0^\mu,-1,0)$ in embedding space. In this notation, the integration measure becomes
\[\int\frac{{\rm d}^{D}y_k}{i\pi^{D/2}}\sim\int\frac{{\rm d}^{D+2}Y_k}{i\pi^{D/2}(Y_k,I_\infty)^D \, \text{vol}(GL(1))}\delta\left(\frac{1}{2}(Y_k,Y_k)\right)\]
with $D=d-2\epsilon$. Therefore, by rewriting $(y_k-x_i)^2-m_i^2=(Y_k,X_k)/(Y_k,I_\infty)$ and $(y_i-y_j)^2=(Y_i,Y_j)/((Y_i,I_\infty)(Y_j,I_\infty))$, we linearize the propagators for a planar integral formally. Once there are exactly $d$ propagators in the integrand involving $Y_k$ for each $k$,  factors $(Y_k,I_\infty)$ will be canceled out in the limit $\epsilon\to 0$, and the integral retains dual conformality in this special case. But for general cases, factors $(Y_k,I_\infty)$ are left. Since $(Y_k,I_\infty)\to0$ accounts for the singularity $y_k^2\to\infty$, we will view $(Y_k,I_\infty)$ as propagators when it appears in the denominator, and consider on-shell condition $(Y,I_\infty)=0$ in our analysis as well.

Now given any family of Feynman integrals, we start with all possible maximal cuts where $d$ degrees of freedom of each loop variable are completely determined\footnote{Note that maximal cuts here for Schubert analysis are different from those usually considered in literature \cite{Frellesvig:2017aai,Bosma:2017ens,Dlapa:2021qsl}, in which the term ``maximal cut" just means to send all propagators of one integral to zero. In our discussion,  maximal cut conditions contain extra ones like $(Y,I_\infty)=0$ furthermore, which always fix all degrees of freedom for $L$ loop momenta.}. Let us collectively denote external dual points as $X_i$ and loop variable as $Y_a$, all of which live in $\mathbb{R}^{d,2}$. For example, the one-loop $d$-gon integral has $d$ poles from propagators of the form $(Y, X_i)$ for $i=1,\cdots d$ and in general also the pole $(Y, I_\infty)$ which we can cut as well. For each independent maximal cut, we denote the solutions as $\{Y\}_\alpha$ with $\alpha$ labelling different solutions, {\it e.g.} there are two cut solutions to $(Y, X_i)=0$ for $i=1,2, \cdots, d$ (note that $Y$ is defined projectively and satisfies $(Y, Y)=0$ thus it is completely fixed by $d$ cut conditions), which we call $Y_{\pm}$. Obviously it is natural to consider Lorentz products of cut solutions; for example, the inverse of the maximal cut (leading singularity) of the $d$-gon, is proportional to the square root, $\Delta=(Y_+, Y_-)$. 

More generally, we can build {\it cross-ratios} from solutions of {\it different} Schubert problems, {\it e.g.} if there are solutions to two different problems which we denote as $Y_{\pm}^1$ and $Y_{\pm}^2$, it is natural to write cross-ratio of the form
\begin{equation}
u(Y^{1,2}_\pm):=\frac{(Y^1_+,  Y^2_+)(Y^1_-, Y^2_-)}{ (Y^1_+, Y^2_-) (Y^1_-, Y^2_+) }.    
\end{equation} 
Indeed we will see that considerations along this line suffice for one-loop symbol letters: in the canonical differential equations (CDE) after we choose uniform transcendental (UT) integral basis \cite{Henn:2013pwa,Henn:2014qga}, the coefficient of the total differential of even $d$-gon with respect to any $(d{-}1)$-gon (also defined near $d$ dimensions) is exactly given by ${\rm d} \log$ of such a letter. Schematically
\begin{equation}
d I_{d}=\sum {\rm d} \log u(Y_\pm^{1,2}) I_{d{-}1} +  \cdots  
\end{equation}
where $1$ denotes the $d$-gon cut and $2$ denotes the corresponding $(d{-}1)$-gon cut with ``infinity" cut as well, {\it i.e.} $(Y, I_{\infty})=0$; eclipse denote other terms in the total differential of $d$-gon. We will see that the full canonical differential equations of one-loop family \cite{Abreu:2017mtm,Arkani-Hamed:2017ahv,Bourjaily:2019exo,Chen:2022fyw,Jiang:2023qnl} can be re-derived in this way, where the symbol letters, or entries of the CDE are given by cross-ratios built from various maximal cut solutions. 

Moving to higher loops, we again consider various maximal cut solutions, not only for integrals in subsectors but also for those lower-loop ones as well. For example, we will consider two-loop UT integrals with two loop variables $Y$ and $Z$. We need maximal cut conditions such as $(Y, X_i)=0$, $(Z, X_i)=0$ (including $I_{\infty}$) and $(Y,Z)=0$, and also one-loop cuts for $Y$ or $Z$. We will illustrate the application of our method using sunrise with two masses in $2-2 \epsilon$ dimensions (which can be extended to the elliptic case with three masses in a similar way~\cite{Bogner:2019lfa,Wilhelm:2022wow}), as well as highly non-trivial cases such as four mass double-box integral family in $4-2 \epsilon$ dimensions. For simplicity, we have chosen this second class of examples near $d=4$ without internal masses (for which our more refined Schubert analysis reveals new structures unseen in~\cite{He:2022tph}). We will see that not only all their symbol letters are obtained from such cross-ratios, but we also see where such letters appear in their CDE. Similar to the dependence of $d$-gon with respect to $(d{-}1)$-gon, we will see that precisely which cross-ratios are needed for the coefficients of ${\rm d} I$ with respect to $I'$, for these two-loop families. 

Some comments are in order. Since the integrals we consider are in general not conformal, such cross-ratios almost always depend on $I_\infty$, which breaks conformal symmetry in embedding space; we also do not fix the normalization of diagonal entries of CDE which depend on the normalization of the UT integral in any case. Moreover, there are numerous maximal cuts and cross-ratios one can construct for a family of integrals, and currently we have not understood precisely which of them do appear in the final answer. However, using these one-loop and two-loop examples, we will find some ``selection rules" which indicates that only certain invariants/cross-ratios should be considered. Relatedly, since we construct our letters using maximal cut solutions, they are also naturally related to similar considerations in Baikov representation of these integrals and maximal cuts~\cite{Baikov:1996iu,Frellesvig:2017aai,Bosma:2017ens}; in particular, we will see that these one-loop cross-ratios are precisely those expressions involving Gram determinants obtained from Baikov representation~\cite{Chen:2022fyw,Jiang:2023qnl}. Although we have chosen to illustrate our method using examples up to two loops, we expect it to work for general Feynman integral families with general masses, at least for planar cases. In the outlook part we will briefly comment on how to go back to the original Schubert analysis for integrals near $d=4$ or $d=3$, as well as extensions to non-planar integrals and those involving elliptic curves.


\section{One-loop integrals and Schubert analysis in embedding space}
As mentioned in introduction, originally Schubert analysis was performed in $d=4$ with momentum twistor variables, and we present a detailed review in appendix \ref{schubert}. In this section, we show how to perform Schubert analysis in general $(d{+}2)$-dimensional embedding space, and we will use the example of one-loop Feynman integrals in arbitrary $d$-dimension with massive propagators as illustration. We will firstly review basic facts about one-loop UT master integrals in general $D=d-2\epsilon$ dimension through embedding language, which naturally inspire us to define their Schubert problems in general $d$. After constructing Lorentzian invariants from their solutions, we will see that one-loop letters can be fully recovered, and our results accord with previous computation precisely.  We end this section by computing some explicit examples.

\subsection{Differential equations for one-loop $n$-gon integrals}
\paragraph{Definitions and notations} Let us first introduce definitions and notations for one-loop $n$-gon UT integrals throughout this section. We will frequently encounter the Gram matrix
\begin{equation}\label{eq:grammatrix}
    G(X_{1},X_{2},\ldots,X_{n})\equiv
    \left(\begin{array}{cccc} (X_1,X_1) & (X_1, X_2) & \cdots & (X_1,X_n) \\ (X_2,X_1) & (X_2,X_2) & \cdots & (X_2,X_n) \\ \vdots & \vdots & \ddots & \vdots \\ (X_n, X_1) & (X_n, X_2) & \cdots & (X_n,X_n) \end{array}\right)
\end{equation}
and use the shorthand notation for determinants $\mathcal{G}=\det G(X_1,\cdots,X_n,I_\infty)$, $\mathcal{G}^0_0=\det G(X_1,\cdots,X_n)$, $\mathcal{G}_{i,0}^{i,0}=\det G(X_1,\cdots,\hat{X_i},\cdots,X_n)$ where the label $0$ denotes the row/column for the infinity point $I_\infty$. We will also use the notation 
\begin{equation}
    G\biggl(\begin{matrix}V_{1},&V_{2},&\ldots,&V_{n}\\U_{1},&U_{2},&\ldots,&U_{n}\end{matrix}\biggr)\equiv
    \left(\begin{array}{cccc} (V_1,U_1) & (V_1,U_2) & \cdots & (V_1,U_n) \\ (V_2,U_1) & (V_2,U_2) & \cdots & (V_2,U_n) \\ \vdots & \vdots & \ddots & \vdots \\ (V_n,U_1) & (V_n,U_2) & \cdots & (V_n,U_n) \end{array}\right) \, .
\end{equation}
and denote determinants of the form
\begin{equation}
\mathcal{G}_A^B=\det G\biggl(\begin{matrix}\hat{X}_B\\\hat{X}_A\end{matrix}\biggr)
\end{equation}
where $\hat{X}_A$ means the sequence $\{X_i\}_{i\in\{1,\cdots,n,0\}\backslash\{A\}}$. 

In practice, we prefer to choose UT basis as master integrals, which read pure, uniform weight-$k$ MPL functions at k-th order of $\epsilon$ expansion, and make the differential equation in canonical form. There are multiple ways to search UT basis \cite{Argeri:2014qva,Gehrmann:2014bfa,Lee:2017oca,Dlapa:2020cwj,Gituliar:2017vzm,Prausa:2017ltv,Meyer:2017joq,Lee:2020zfb,Chen:2022lzr}, and in this paper we adopt the definition of n-gon UT integrals in \cite{Chen:2022lzr} as the following
\begin{equation}
    \begin{aligned}
        I[1,2,\ldots,n;D]=\epsilon^{\lceil \frac{n}{2} \rceil}\int\frac{\mathrm{d}^{D}y}{i\pi^{D/2}}\frac{\mathcal{N}_{n}}{D_{1}D_{2}\ldots D_{n}}
    \end{aligned} \ ,
\end{equation}
where $\epsilon^{\lceil \frac{n}{2} \rceil}$ unifies the transcendental weight of these integrals; the spacetime dimension $D=n-2\epsilon$ for even $n$ and $D=n{+}1-2\epsilon$ for odd $n$, $D_i=(y-x_i)^2-m_i^2$ denote the propagators, and the numerator $\mathcal{N}_{n}$ is defined as
\begin{equation}\label{eq:Ln}
    \begin{aligned}
        \mathcal{N}_{n}=\left\{\begin{array}{cc}
            \sqrt{\frac1{(-2i)^n}\det G(X_1,\cdots,X_n)} & , \, n \text{ is even} \\
            \sqrt{-\frac1{(-2i)^{n-1}}\det G(X_1,\cdots,X_n,I_\infty)} & , \, n \text{ is odd}
        \end{array}\right. \,
    \end{aligned}
\end{equation}
where the imaginary unit $i$ in the normalization is used to keep the coefficients real in CDE. Rewriting the integrals in embedding formalism, we have
\begin{equation}\label{embedintegrand}
    \begin{aligned}
        I[{1}, & {2},\ldots,{n};D]=\\
       & \left\{\begin{array}{cc}
        \epsilon^{\lceil \frac{n}{2} \rceil}\int\frac{\mathrm{d}^{D{+}2}Y}{i\pi^{D/2}\text{vol}(GL(1))}\delta(\frac{1}{2}(Y,Y))\frac{\mathcal{N}_{n}(Y,I_\infty)^{2\epsilon}}{(Y,X_1)\cdots(Y,X_n)} & n \text{ is even},\ D=n-2\epsilon \\
        \epsilon^{\lceil \frac{n}{2} \rceil}\int\frac{\mathrm{d}^{D{+}2}Y}{i\pi^{D/2}\text{vol}(GL(1))}\delta(\frac{1}{2}(Y,Y))\frac{\mathcal{N}_{n}(Y,I_\infty)^{2\epsilon}}{(Y,X_1)\cdots(Y,X_n)(Y,I_\infty)} & n \text{ is odd},\ D=n+1-2\epsilon
    \end{array}\right. \, .
    \end{aligned}
\end{equation}
Differential equations of one-loop integrals and their alphabet have been thoroughly computed in \cite{Abreu:2017mtm,Caron-Huot:2021xqj,Chen:2022fyw,Jiang:2023qnl}, and in the following part of this section we will use Schubert analysis in embedding space to rederive them. Note that the one-loop UT integrals are defined in $D=d-2\epsilon$ with $d$ being an integer, and by definition Schubert analysis is performed in $d$ dimensions. It is very interesting that such analysis allows us to reconstruct the full alphabet in $D$ dimensions, as we are familar in $d=4$ case \cite{He:2022tph} and we will see more evidence for general $d$ now.

\paragraph{One-loop Schubert analysis}
Generally speaking, the Schubert problem associated with a particular integral is closely related to solving its leading singularity. Following on-shell conditions by sending all of its propagators to zero, we focus on the case where loop momenta are completely determined as maximal cut solutions, and show that symbol letters can be obtained from conformal invariants of these solutions. For one-loop integrals in embedding space, it is natural to define the one-loop $d$-dimensional Schubert problems for even $n$-gon when $n=d$ and for odd $n$-gon when $n=d{-}1$. Solutions $ Y $ for Schubert problems satisfy $d$ on-shell conditions
\begin{equation}\left\{\begin{array}{cc}
       &(Y,X_1)=\cdots=(Y,X_n)=0\ \   n=d \text{ is even} \\
       &(Y,X_1)=\cdots=(Y,X_{n})=(Y,I_\infty)=0\ \ \   n=d{-}1 \text{ is odd}
   \end{array}\right. \, ,
   \end{equation}
as well as projectivity and null condition $(Y,Y)=0$, which together fix $(d{+}2)$ degrees of freedom of $Y$. Due to the quadratic condition $(Y,Y)=0$, we have two solutions $Y_+$ and $Y_-$ for each problem. It can be checked that the discriminant of the quadratic condition is proportional to $\sqrt{\mathcal{G}^0_0}$ for an even $n$-gon and proportional to $\sqrt{\mathcal{G}}$ for an odd $n$-gon. In another word, if we consider the $d$-dimensional leading singularities of the integrand
\[\frac{1}{(Y,X_1)\cdots(Y,X_n)},\ \  \frac{1}{(Y,X_1)\cdots(Y,X_n)(Y,I_\infty)}\]
for even or odd $n$ respectively, we will get $LS\propto\frac1{\sqrt{\mathcal{G}^0_0}}$ for even $n$, and $LS\propto\frac1{\sqrt{\mathcal{G}}}$ for odd $n$, which explains the normalization factors in \eqref{eq:Ln}. These maximal cut solutions, $Y_\pm$, of various Schubert problems will become the most basic ingredients to reconstruct one-loop alphabet, as we will see.


It is well known that the total differential of an $n$-gon ${\rm d} I[1,\cdots,n;D]$ can be expressed by a linear combination of $n$-gon $ I=I[1,\cdots,n;D] $ , $(n{-}1)$-gon $I^i=I[1,\cdots,\hat{i},\cdots,n;D]$ and $(n{-}2)$-gon $I^{i,j}{=}I[1,\cdots,\hat{i},\cdots,\hat{j},\cdots,n;D]$, with coefficients being some $ {\rm d} \log$ one-form:
\begin{equation} \label{eq:oneloopCDE}
    {\rm d} I = \epsilon \left( x \,  {\rm d} \log (W) \, I + y \sum_i {\rm d} \log(W_i) \, I^i + z \sum_{i<j} {\rm d} \log (W_{i,j}) \, I^{i,j} \right) ,
\end{equation}
where $ \{x,y,z\} $ are some rational numbers and $ \{W, W_i, W_{i,j}\} $ are known as the symbol letters. Let us recover these letters from Schubert analysis recursively. For an arbitrary even $n$-gon (in $D=n-2\epsilon$) and its Schubert solutions $Y_\pm$, $\{X_1,\cdots,X_n,Y_+,Y_-\}$ forms a complete basis for $(d{+}2)$-dimensional embedding space. For arbitrary $Z$ in the space, we have the completeness relation \cite{Herrmann:2019upk}
\begin{equation}\label{complete}
    (Z,Z)=2\frac{(Z,Y_+)(Z,Y_-)}{(Y_+,Y_-)}+\sum_{a,b=1}^d (G^{-1})_{a,b}(Z,X_a)(Z,X_b) 
\end{equation}
where $G=G(X_1,\cdots,X_n)$. Plugging in $Z=I_\infty$, product of the two solutions respect the relation
\begin{equation}
\Delta = (Y_+,Y_-)\propto \frac{-2}{\left(\sum_{a,b=1}^n (G^{-1})_{a,b}\right)}=
       \frac{2 \mathcal{G}^0_0}{\mathcal{G}}
\end{equation}
On the other hand, for the odd $n$-gon in $n{+}1$ dimensions, although completeness condition is no longer applicable due to the on-shell condition  $(Y_\pm,I_\infty)=0$, we can still directly solve the Schubert problem and derive that  
\begin{equation}
\Delta = (Y_+,Y_-)\propto
       -\frac{\mathcal{G}}{2 \mathcal{G}^0_0}
\end{equation}
In both cases we see that $ \Delta $ actually accounts for the letter $ W $ in  
\eqref{eq:oneloopCDE}, {\it i.e.} ${\rm d} \log (W) \propto {\rm d} \log (\Delta)$ . It is not surprising, since the diagonal element should stem purely from the Schubert problem of $n$-gon itself, and $\Delta$  is the only Lorentzian invariant we can construct from an individual Schubert problem\footnote{Note that $\Delta$ always has ambiguities. For instance, changing fixing of projectivity from $(Y_\pm,I_\infty)=1$ to $(Y_\pm,I_\infty)=k$ will rescale $\Delta\to k^2\Delta$, and only cross-ratios of the solutions remain invariant. However, even for differential equations themselves we know that it is free for us to change normalization of master integrals $I\to (x)^\epsilon I$, where $x$ is any combination of kinematics variables. Such a transformation will also changes the diagonal letters of DE but leaves the off-diagonal letters invariant. So we retain the ``$\propto$" sign in our discussion for these letters as well. It should be clarified that in all examples we discuss in this paper, we can still consider all cross-ratios containing products $(Y_+,Y_-)$, from which the multiplicative space of all rational letters can be reproduced invariantly.}.     

Now we turn to the dependence of $n$-gon to lower-gons. Let us first consider even $n$-gon in $n$ dimensions. In this case, dependence of $dI$ on $(n{-}1)$-gon $I^i$ should be parity odd under the transformation  $\sqrt{\mathcal{G}^0_0}\to -\sqrt{\mathcal{G}^0_0}$ and $\sqrt{\mathcal{G}^i_i}\to -\sqrt{\mathcal{G}_i^i}$, thus ${\rm d} \log W_i \to -{\rm d} \log W_i $ under each of the two transformations, since integrals $I$ or $I^i$ are transformed to $-I$ or $-I^j$ respectively. In terms of Schubert solutions $\{Y_\pm, Y^i_\pm\}$, where $Y_\pm$ are the solutions for the $n$-gon Schubert problems and $Y_\pm^i$ are those for $(n{-}1)$-gon $ I^i $, the only reasonable candidate is then
\begin{equation}\label{type1}
u(\{Y_\pm, Y^i_\pm\})=\frac{(Y_+,Y_+^i)(Y_-,Y_-^i)}{(Y_+,Y_-^i)(Y_-,Y_+^i)}.
\end{equation}
It is easy to see that \eqref{type1} naturally enjoys the correct parity property, and therefore is inferred to be the coefficient in front of $(n{-}1)$-gon, \textit{i.e.} $ {\rm d} \log(W_i) \propto {\rm d} \log(u(\{Y_\pm, Y^i_\pm\})) $. In addition, \eqref{type1} can be written in Gram determinants explicitly as
\begin{equation}\label{type1p}
    \biggl(\frac{\mathcal{G}^i_0+\sqrt{\mathcal{G}^i_i\mathcal{G}^0_0}}{\mathcal{G}^i_0-\sqrt{\mathcal{G}^i_i\mathcal{G}^0_0}}\biggr)^2 .
\end{equation} 

Finally, let us turn to the elements between $n$-gon and $(n{-}2)$-gon. Similar to \eqref{type1p}, dependence of ${\rm d}I$ on $I^{i,j}$ should be parity odd under transformations $\sqrt{\mathcal{G}^{i,j,0}_{i,j,0}}\to -\sqrt{\mathcal{G}_{i,j,0}^{i,j,0}}$ and $\sqrt{\mathcal{G}^0_0}\to -\sqrt{\mathcal{G}^0_0}$.  Problems immediately arise; two Schubert problems associated to $n$-gon and $(n{-}2)$-gon live in different dimension, therefore cannot be combined like \eqref{type1} directly. 
Nevertheless, from dimensional shifting relation, we know that even $n$-gon in $D=n-2\epsilon$ is related to that in $D=n{-}2{-}2\epsilon$ by $I^{(D)}_n\propto \frac{\mathcal{G}^0_0}{\mathcal{G}} I^{(D-2)}_n + \text{lower point integrals}$, and in $D=n-2-2\epsilon$ the $n$-gon integral contains many different $(n{-}2)$-gon sub-topologies. Each of them is associated to a $(n-2)$-dimensional Schubert problem, so that we can combining their solutions to construct more letters. For instance, for each $(n{-}2)$-gon sub-topology $I^{i,j}$ in the differential equation, we can choose another arbitrary $k\neq i,j$ such that $I^{i,k}$ and $I^{j,k}$ are two $(n{-}2)$-gons. Now we have a ``$6$-point configuration"
\begin{equation}\label{6pt}
    \{Y_+^{i,j},Y_-^{i,j},Y_+^{i,k},Y_-^{i,k},Y_+^{j,k},Y_-^{j,k}\}
\end{equation}
We can construct $9$ multiplicative independent cross-ratios from this configuration \footnote{This configuration in $d=4$ was used for predicting third entries of $12$-point double-box in \cite{Morales:2022csr}. We give more details about this configuration in Appendix \ref{schubert}}. Among all the letters, we pick up a special combination
\begin{equation}\label{type2}
\frac{(Y^{i,j}_+,Y^{i,k}_+)(Y^{i,j}_+,Y^{i,k}_-)(Y^{i,j}_-,Y^{j,k}_+)(Y^{i,j}_-,Y^{j,k}_-)}{(Y^{i,j}_-,Y^{i,k}_+)(Y^{i,j}_-,Y^{i,k}_-)(Y^{i,j}_+,Y^{j,k}_+)(Y^{i,j}_+,Y^{j,k}_-)}
\end{equation}
Although it looks a little bit intricate, this combination turns out to be reasonable for our need. Firstly, the letter acts parity odd when transforming $\sqrt{\mathcal{G}^{i,j,0}_{i,j,0}}\to -\sqrt{\mathcal{G}_{i,j,0}^{i,j,0}}$ by definition. 
Secondly, secretly this letter is also parity odd under the transformation $\sqrt{\mathcal{G}^0_0}\to -\sqrt{\mathcal{G}^0_0}$. To see this point, following the completeness relation \eqref{complete} we have\footnote{We thank Yichao Tang for suggesting this proof.}
\begin{align}
   &\log\frac{(Y^{i,j}_+,Y^{i,k}_+)(Y^{i,j}_+,Y^{i,k}_-)(Y^{i,j}_-,Y^{j,k}_+)(Y^{i,j}_-,Y^{j,k}_-)}{(Y^{i,j}_-,Y^{i,k}_+)(Y^{i,j}_-,Y^{i,k}_-)(Y^{i,j}_+,Y^{j,k}_+)(Y^{i,j}_+,Y^{j,k}_-)} =\int^{Y^{i,j}_+}_{Y^{i,j}_-}{\rm d}\log\frac{(Y,Y^{i,k}_+)(Y,Y^{i,k}_-)}{(Y,Y^{j,k}_+)(Y,Y^{j,k}_-)}\nonumber\\
   &=\int^{Y^{i,j}_+}_{Y^{i,j}_-}{\rm d}\log\frac{(Y^{i,k}_+,Y^{i,k}_-)(Y,X_j)^2}{(Y^{j,k}_+,Y^{j,k}_-)(Y,X_i)^2}{=}\log\left(\frac{(Y^{i,j}_+,X_j)(Y^{i,j}_-,X_i)}{(Y^{i,j}_-,X_j)(Y^{i,j}_+,X_i)}\right)^2{=}\log\left(\frac{\mathcal{G}^{i,0}_{j,0}+\sqrt{-\mathcal{G}_{i,j,0}^{i,j,0}\mathcal{G}^0_0}}{\mathcal{G}^{i,0}_{j,0}-\sqrt{-\mathcal{G}_{i,j,0}^{i,j,0}\mathcal{G}^0_0}}\right)^2
\end{align}
One sees that \eqref{type2} is actually independent of $k$ we choose, and enjoy the correct parity property we need.  Therefore, \eqref{type2} are the ideal candidates accounting for symbol letters between $n$-gon and $(n{-}2)$-gon. 

To make a comparison, since at $n$-dimension we have $(X_i,Y_\pm)=(X_j,Y_\pm)=0$, we can uplift above cross-ratio from $n{-}2$ dimension to $n$-dimension by simply introducing a secondary Schubert problem as \[(Y,X_a)=(Y,Y_+)=(Y,Y_-)=0\]
with $a\neq i,j$ and $Y_\pm$ being the solutions of Schubert problem for $n$-gon. Denoting its solutions as $Y^{(ij)}_\pm$, the letter above is equal to the square of
\begin{equation}\label{type2p}
\frac{(Y^{(ij)}_+,X_i)(Y^{(ij)}_-,X_j)}{(Y^{(ij)}_+,X_j)(Y^{(ij)}_-,X_i)}\,
\end{equation}
This result was also presented in \cite{Herrmann:2019upk}. 

For the total differential of odd $n$-gon, things are almost the same. Note that besides $(Y_+,Y_-)$ gives the diagonal element, coefficients between $n$-gon and $(n{-}1)$-gon or $(n{-}2)$-gon are both of the second type like \eqref{type2}, since Schubert problems for $n$-gon are in $(n{+}1)$-dimension, while $(n{-}1)$-gon and $(n{-}2)$-gon are in $(n{-}1)$-dimension. Therefore, we can correspondingly construct
\begin{equation}
  \frac{(Y^{(i\infty)}_+,X_i)(Y^{(i\infty)}_-,I_\infty)}{(Y^{(i\infty)}_-,X_i)(Y^{(i\infty)}_+,I_\infty)},\ \frac{(Y^{(ij)}_+,X_i)(Y^{(ij)}_-,X_j)}{(Y^{(ij)}_+,X_j)(Y^{(ij)}_-,X_i)} 
\end{equation}
for $(n{-}1)$-gon and $(n{-}2)$-gon respectively, {\it i.e.}, we regard $n$-gon as a $(n{+}1)$-gon containing $(Y,I_\infty)$ as its propagator; $(n{-}2)$-gons (or $(n{-}1)$-gon) are then its $n$-gon sub-topologies, with (or without) $(Y,I_\infty)$ as propagator. Rewritten in Gram determinants, the two letters read
\begin{equation}
\frac{\mathcal{G}^{i}_{0}+\sqrt{-\mathcal{G}_{i,0}^{i,0}\mathcal{G}}}{\mathcal{G}^{i}_{0}-\sqrt{-\mathcal{G}_{i,0}^{i,0}\mathcal{G}}},\ \ \frac{\mathcal{G}^{i}_{j}+\sqrt{-\mathcal{G}_{i,j}^{i,j}\mathcal{G}}}{\mathcal{G}^{i}_{j}-\sqrt{-\mathcal{G}_{i,j}^{i,j}\mathcal{G}}}
\end{equation}
up to a minus sign.

As a summary, differential equations for one-loop integrals in $n$-dimension are then determined to be 
\begin{equation}\label{DES1}
   \begin{aligned}
       \mathrm{d} I=\epsilon \left( x_1\  \mathrm{d}\log\frac{2 \mathcal{G}^0_0}{\mathcal{G}} I+ y_1 \sum_{i}\mathrm{d}\log \frac{\mathcal{G}^i_0+\sqrt{\mathcal{G}^i_i\mathcal{G}^0_0}}{\mathcal{G}^i_0-\sqrt{\mathcal{G}^i_i\mathcal{G}^0_0}}I^{i} \right.
       \left.+z_1\sum_{i<j}\mathrm{d}\log\frac{\mathcal{G}^{i,0}_{j,0}+\sqrt{-\mathcal{G}_{i,j,0}^{i,j,0}\mathcal{G}^0_0}}{\mathcal{G}^{i,0}_{j,0}-\sqrt{-\mathcal{G}_{i,j,0}^{i,j,0}\mathcal{G}^0_0}}I^{i,j}\right) \, ,
   \end{aligned}
\end{equation}
for I is an even $n$-gon, and
\begin{equation}\label{DES2}
   \begin{aligned}
     \mathrm{d} I=\epsilon \left( x_2\ \mathrm{d}\log\frac{ 2\mathcal{G}^0_0}{\mathcal{G}} I+ y_2 \sum_{i}\mathrm{d}\log \frac{\mathcal{G}^{i}_{0}+\sqrt{-\mathcal{G}_{i,0}^{i,0}\mathcal{G}}}{\mathcal{G}^{i}_{0}-\sqrt{-\mathcal{G}_{i,0}^{i,0}\mathcal{G}}}I^{i} \right. 
       \left.+z_2\sum_{i<j}\mathrm{d}\log\frac{\mathcal{G}^{i}_{j}+\sqrt{-\mathcal{G}_{i,j}^{i,j}\mathcal{G}}}{\mathcal{G}^{i}_{j}-\sqrt{-\mathcal{G}_{i,j}^{i,j}\mathcal{G}}}I^{i,j}\right) \, ,
   \end{aligned}
\end{equation}
for $I$ is an odd $(n{-}1)$-gon. $\{x_i,y_i,z_i\}$ in the expression are certain rational numbers that need to be fixed later on.

\paragraph{Comparison with CDE} The one-loop differential equations and the letters has been investigated from various aspects, such as from direct computation \cite{Henn:2022ydo}, diagrammatic coaction \cite{Abreu:2017enx,Abreu:2017mtm}, hyperbolic geometries \cite{Bourjaily:2019exo}, dual forms \cite{Caron-Huot:2021xqj}, $\mathcal{A}$-determinants \cite{Dlapa:2023cvx}, and also from Baikov ${\rm d\log}$ representations \cite{Chen:2022fyw,Jiang:2023qnl} and intersection theory \cite{Chen:2023kgw},{\it etc.}. There are also papers that focus on the finite truncation of one-loop DE \cite{Spradlin:2011wp,Caron-Huot:2014lda,Arkani-Hamed:2017ahv,Herrmann:2019upk}. Our expressions \eqref{DES1} and \eqref{DES2} from Schubert analysis show perfect agreement with the former result, once we set $\{x_i,y_i,z_i\}=\{-1,\frac12,\frac14\}$. More details about Baikov representations and fixing rational numbers can be found appendix \ref{Baikov}.

\subsection{Explicit examples of one-loop integrals}
Let us illustrate the general results above with some explicit examples of differential equations for one-loop integrals. 

\paragraph{$d=4$ massless boxes with massive propagators}
Let us firstly explicitly present the differential equations for boxes at $D=4-2\epsilon$. The fully massive box integral, {\it i.e.} both internal legs and external legs are massive, has been discussed in \cite{Bourjaily:2019exo} by relating to hyperbolic volume. In this section, on the other hand, we will mainly focus on their symbol letters. To simplify the problem, we consider the case when external legs become massless while masses of the propagators vary. It depends on $5$ ratios of six variables $\{s,t,m_i^2\}_{i=1\cdots 4}$. When $m_i^2=m^2$ for all $i$, the integral degenerates to the case discussed in \cite{Caron-Huot:2014lda}. 

Let us work out one of the entries more explicitly as an illustration for our procedure. Suppose the two Schubert solutions for the four-mass boxes $(Y,X_1)=\cdots=(Y,X_4)=(Y,Y)=0$ reads $Y_+$ and $Y_-$, then we consider Schubert problem for triangle $I^{4}$, whose Schubert problem reads
\begin{equation}\label{Schtri}
(Y,X_1)=(Y,X_2)=(Y,X_3)=(Y,I_\infty)=(Y,Y)=0
\end{equation}
in $(4+2)$-dimensional embedding space. To solve the problem, without loss of generality, we parametrize $Y^4_\pm=\sum_{i=1}^4a_{i,\pm} X_i+a_{p,\pm}Y_++a_{m,\pm}Y_-$, since $\{X_i\}_{i=1,\cdots 4} \bigcup \{ Y_{+}, Y_{-} \} $ in fact form a complete basis in the embedding space. Letter \eqref{type1} written in these undetermined coefficients is simply $\frac{a_{m,+}a_{p,-}}{a_{p,+}a_{m,-}}$ since $(X_i,Y_\pm)=0$. Fixing $a_4=1$ by projectivity and applying the parametrization to \eqref{Schtri}, the solutions for the coefficients then read
\begin{align}
    &(a_{1,\pm},a_{2,\pm},a_{3,\pm})^T=G(X_1,X_2,X_3)^{-1}\cdot((X_4,X_1),(X_4,X_2),(X_4,X_3))^T\nonumber\\
    &a_{m,\pm}=-(1+a_{1,\pm}+a_{2,\pm}+a_{3,\pm}+a_{p,\pm})\nonumber\\
    &-2a_{p,\pm}^2(Y_+,Y_-)-2a_{p,\pm}(1+a_1+a_2+a_3)(Y_+,Y_-)\nonumber\\
    &+(a_1,a_2,a_3,1)\cdot G(X_1,X_2,X_3,X_4)\cdot(a_1,a_2,a_3,1)^T=0
\end{align}
Discirminant of the last equation, as can be checked, is proportional to $\sqrt{\mathcal{G}_i^i\mathcal{G}^0_0}$ as we expect. Following the relation $a_{p,+}+a_{p,-}=-(1+a_1+a_2+a_3)=a_{m,+}+a_{p,+}$, ratio $\frac{a_{m,+}a_{p,-}}{a_{p,+}a_{m,-}}$ is then naturally a square of certain odd letter $(\frac{a_{p,-}}{a_{p,+}})^2$, whose explicit form is \eqref{type1p}.

After letters in \eqref{DES1} and \eqref{DES2} are thoroughly computed, it is then straightforward to write down the differential equations for the system. For example, total differential of the box can be expressed by triangles and bubbles, whose explicit form reads
\begin{equation}\label{example1}
    {\rm d}I=\epsilon\left(-{\rm d}\log \frac{2r_0}{R_0} I+\frac12\sum_{i=1}^4{\rm d}\log\frac{S_i+\sqrt{r_0r_i}}{S_i-\sqrt{r_0r_i}}I^{i}+\frac14\sum_{0<i<j\leq 4}{\rm d}\log\frac{T_{i,j}+\sqrt{r_0r_{i,j}}}{T_{i,j}-\sqrt{r_0r_{i,j}}}I^{i,j}\right)
\end{equation}
with the shorthand ($v_1=v_3=s$, $v_2=v_4=t$)
\begin{align}
&r_0=\mathcal{G}_0^0,\ \ R_0=2s t(s+t)\\
&r_i=m_i^2m_{i+1}^2{+}m_{i+1}^2m_{i+2}^2{-}m_i^2m_{i+1}^2{-}m_{i{+}1}^4{-}m_{i{+}1}^2v_i\\
&r_{i,i+2}=(m_i^2{+}m_{i+2}^2{-}v_i)^2{-}4m_i^2m_{i+2}^2\\
&S_i=v_{i+1}(v_{i+1}(m_i^2-m_{i+2}^2-v_i)+v_i(m_{i+1}^2-2m_{i+2}^2+m_{i+3}^2))\\
&T_{i,j}=v_{i}(m^2_{j+2}(m^2_j-m^2_{j+2}+v_j)-m^2_i(m^2_{j}-m^2_{j+2}-v_j))+v_j (m^2_i-m^2_{i+2})(m^2_j-m^2_{j+2})
\end{align}
Since all these integrals are both ultraviolet and infrared convergent, we can send $\epsilon\to0$ in the differential equations and reduce the DE system to purely $D=4$. This amounts to rescale the basis as (the rescaling is to compensate the normalization factor $\lceil \frac{n}{2} \rceil$ we introduce in the definition)
\[(\mathcal{I},\mathcal{I}^{i},\mathcal{I}^{i,i+2})=\left(\frac1{\epsilon^2}I,\frac1{\epsilon^2}I^{i},\frac1{\epsilon}I^{i,i+2}\right)\]
and take the limit $\epsilon\to0$. After the procedure, dependence of $I^{i}$ in the expression completely drops out, and we come back to the famous formula \cite{Spradlin:2011wp}
\begin{equation}
    {\rm d}\mathcal{I}=\frac14\sum_{0<i<j\leq 4}{\rm d}\log\frac{T_{i,j}+\sqrt{r_0 r_{i,j}}}{T_{i,j}-\sqrt{r_0 r_{i,j}}} \mathcal{I}^{i,j}
\end{equation}

\paragraph{$6d$ massless hexagon} Differential equations for one-loop massless (without external and internal masses) $d$-dimensional hexagon have been presented in \cite{Henn:2023vbd}. The alphabet consists of $93$+$10$ symbol letters, $93$ of which are from union of alphabets for one-mass pentagon sub-topologies. The other $10$, on the other hand, are genuine $6$-point letters and of interests. They come from the total differential of hexagon integral, which can be explicitly written down as,
\begin{equation}\label{DEhex}
   \begin{aligned}
       \mathrm{d} I_6 = & \epsilon \left( -\mathrm{d}\log (Y_+,Y_-) I+ \frac{1}{4}\sum_{i=1}^6\mathrm{d}\log\frac{(Y_+,Y_+^i)(Y_-,Y_-^i)}{(Y_+,Y_-^i)(Y_-,Y_+^i)} I^{i} \right. \\
       &\left.+\frac{1}{4}\sum_{1\leq i<j\leq6}\mathrm{d}\log\frac{(Y^{(ij)}_+,X_i)(Y^{(ij)}_-,X_j)}{(Y^{(ij)}_+,X_j)(Y^{(ij)}_-,X_i)}I^{i,j}\right) \, ,\\
       = & \epsilon \left( -\mathrm{d}\log W_{40}\ I+ \frac{1}{2}\sum_{i=1}^6\mathrm{d}\log W_{i+94}\ I^{i}+\frac{1}{4}\sum_{1\leq i\leq 6}\mathrm{d}\log\ w_i\ I^{i,i{+}1}\right.\\
       &\left. -\frac14\sum_{i=1}^6\mathrm{d}\log (w_{i{+}1}w_{i{+}2}) I^{i,i{+}2}-\frac14\sum_{i=1}^3\mathrm{d}\log (w_1 w_2 w_3) I^{i,i{+}3}\right)
   \end{aligned}
\end{equation}
with the notation $\{w_i,w_{i+3}\}_{i=1,\cdots,3}=\{W_{i+100},W_{i+100}\}_{i=1,\cdots,3}$.\eqref{DEhex} exactly matches the first line of the DE computed in \cite{Henn:2023vbd}.  At order $\epsilon^0$, letters $W_{i+101}$ for $i=0,1,2$ read the last entries of $I_6$, and the result goes back to conformally invariant hexagon integral \cite{Spradlin:2011wp}. Keeping on applying investigation to pentagon and lower $n$-gon integrals in this CDE system, finally we can reproduce the full result in \cite{Henn:2023vbd}. 

\subsection{One-loop alphabets and DE near odd dimensions}
In the end of this section we briefly comment on odd-dimensional one-loop alphabet and the differential equations, which is quite similar to the even-dimensional ones. Parallel to the previous case, definitions for the $n$-gon canonical integrals are introduced as followings:
\begin{equation}
    \begin{aligned}
        I[a_{1},a_{2},\ldots,a_{n};D]=\epsilon^{\lceil \frac{n-1}{2} \rceil}\int\frac{\mathrm{d}^{D}y}{i\pi^{D/2}}\frac{\mathcal{N}_{n}}{D_{1}D_{2}\ldots D_{n}}
    \end{aligned}
\end{equation}
with $D=n-2\epsilon$ when $n$ is {\it odd}, and $D=n+1-2\epsilon$ when $n$ is {\it even}. Normalization factors now read 
\begin{equation}\label{eq:Ln2}
    \begin{aligned}
        \mathcal{N}_{n}=\left\{\begin{array}{cc}
            \sqrt{\frac1{(-2i)^n}\det G(X_1,\cdots,X_n)} & n \text{ is odd} \\
            \sqrt{-\frac1{(-2i)^{n-1}}\det G(X_1,\cdots,X_n,I_\infty)} & n \text{ is even}
        \end{array}\right. \, .
    \end{aligned}
\end{equation}
accordingly. We see that the two definitions are actually just a simple exchange of the expressions for even-gon and odd-gon, and then shifting $D\to D+1$ in the original even-dimensional case. Therefore, differential equations for the $n$-gons can be written down by analogy as \eqref{DES1} for I which is an odd-gon, and \eqref{DES2} for $I$ which is an even-gon. $I^i$ (and $I^{i,j}$) are again $n{-}1$-gon (or $n{-}2$-gon) with propagator $X_i$ (or $X_i$ and $X_j$) absent. And similarly, we should define Schubert problems by the on-shell conditions
\begin{equation}\label{eq:cutcondition}
\left\{\begin{array}{cc}
       &(Y,X_1)=\cdots=(Y,X_n)=0\ \   n=d \text{ is odd} \\
       &(Y,X_1)=\cdots=(Y,X_{n})=(Y,I_\infty)=0\ \ \   n=d{-}1 \text{ is even}
   \end{array}\right. \, .
\end{equation}
respectively at $d$-dimension, $D=d-2\epsilon$ again.

As the end of this section, let us explore DE for one-loop box with four internal massive propagators and massless external legs at $D=5-2\epsilon$ again and make a comparison. Like the case for $D=4-2\epsilon$, there are again $11$ independent UT integrals in this sector, which can be chosen as $1$ box in $D=5-2\epsilon$, $4$ one-mass triangles and $2$ bubbles in $D=3-2\epsilon$, and $4$ tadpoles in $D=1-2\epsilon$. Total differential of the box integrals can be expended by linear combinations of the two bubbles and four triangles, and the relation reads\footnote{Careful readers may notice that only two bubbles remain in the differential equation. Other bubbles are reducible, and in this case box won't depend on those bubbles. This can be seen easily by examining the degeneration of general formula \eqref{DES1} and \eqref{DES2} for odd dimension. That is, \eqref{DES1} and \eqref{DES2} always hold in general kinematics configuration. When we take the external legs to be massless as in our example, some bubbles become reducible and the letters before these bubbles degenerate to $\log 1=0$. So we can directly discard corresponding terms.}
\begin{equation}
    {\rm d}I=\epsilon\left(-{\rm d}\log\frac{2r_0}{R_0}I+\frac12\sum_{i=1}^4{\rm d}\log(\frac{U_i{+}\sqrt{R_0R_i}}{U_i{-}\sqrt{R_0R_i}})I^{i}+\frac14\sum_{\{i,j\}=\{1,3\},\{2,4\}}{\rm d}\log(\frac{V_i{+}\sqrt{R_0v_i}}{V_i{-}\sqrt{R_0v_i}})I^{i,j}\right)
\end{equation}
shorthand in the expression are ($r_0=\mathcal{G}^0_0$ and $R_0=2s t(s+t)$ again)
\begin{align}\label{example2}
    &R_i=2v_{i+1}(m_{i+1}^2m_{i+2}^2-m_{i+2}^4-m_{i+1}^2m_{i+3}^2+m_{i+2}^2m_{i+3}^2-m_{i+2}^2v_{i+1})\\
    &U_i=v_{i+1}(v_i(m_{i+1}^2-2m_{i+2}^2+m_{i+3}^2+v_{i+1}(m_i^2-m_{i+2}^2-v_i)))\\
    &V_i=v_i(2v_{i+1}+v_i)
\end{align}
Comparing \eqref{example2} with \eqref{example1}, we see the last two terms are distinct. $\epsilon\to0$ limit in this case leads to a similar expression as
\begin{equation}
    {\rm d}\mathcal{I}=\frac12\sum_{i=1}^4{\rm d}\log(\frac{U_i{+}\sqrt{R_0R_i}}{U_i{-}\sqrt{R_0R_i}})\mathcal{I}^{i}+\frac14\sum_{\{i,j\}=\{1,3\},\{2,4\}}{\rm d}\log(\frac{V_i{+}\sqrt{R_0v_i}}{V_i{-}\sqrt{R_0v_i}})\mathcal{I}^{i,j}
\end{equation}
at $D=5$, where we rescale $\mathcal{I}=\frac1{\epsilon^2}I$, $\mathcal{I}^{i}=\frac1{\epsilon}I^{i}$ and $\mathcal{I}^{i,j}=\frac1{\epsilon}I^{i,j}$.

\section{Applications of Schubert analysis to higher-loop integrals}
Now that we have understood one-loop differential equations and their symbol letters from Schubert analysis, we move to higher loops. Recall that the symbol alphabets for a lot of multi-loop Feynman integrals near $d=4$ dimensions have been generated from Schubert analysis in twistor space. However, it is still highly non-trivial to see if they can be extended to embedding space, and more importantly here we will attempt to obtain not only the full alphabet, but also more detailed information regarding the canonical differential equations for certain two-loop integral families. One new feature of higher-loop integrals is that their master integrals start to contain distinct independent scalar products (ISPs) of $\{y_i,x_i\}$ besides the propagators, leading to a larger integral basis, as well as more intricate mathematical structures. In this section, we take the first step in understanding their structure through Schubert analysis, by exploring where each symbol letter may appear in the differential equations. 

In principle, the Schubert analysis can be extended to general $L$-loop planar integrals in embedding space. Generally speaking, for an $L$-loop integral whose loop momenta are defined in $D=d-2\epsilon$ dimension, we introduce Schubert problem at $d$ dimension for the integral by sending all its propagator on-shell to fix $dL$ degrees of freedom for the loop momenta (with $2L$ additional degrees of freedom in embedding formalism fixed by conditions $(Y_i,Y_i)=0$ and projectivity). Solutions for the loop momenta then become basic ingredients for us to construct symbol letters for the integrals. In this section we will see that higher-loop Schubert analysis is tightly related to individual UT basis in the integral family. Consequently, differential equations for two-loop integrals enjoy similar structures like the one-loop case. We will work with the sunrise integral family in $D=2-2\epsilon$ with two masses, as well as four-mass double-box family in $D=4-2\epsilon$ to illustrate this idea. 

\subsection{Sunrise in $D=2-2\epsilon$}
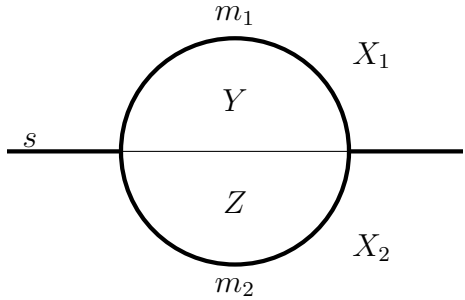
\begin{figure}[htbp]
    \centering
     \begin{tikzpicture}[scale=1.5]
 	\draw[black,ultra thick] (-2,0)--(-1,0);
 	\draw[black,ultra thick] (2,0)--(1,0);
 	\filldraw[color=black,fill=white,ultra thick] (0,0) circle (1);
 	\draw[black] (-1,0)--(1,0);
 	\node (s) at (-1.8,0.1) { $s$};
 	\node (m1) at (0,1.2) { $m_1$};
  \node (Y) at (0,0.45) {$Y$};
    \node (Z) at (0,-0.45) {$Z$};
 	\node (m2) at (0,-1.2) { $m_2$};
  \node (X1) at (1.2,0.85) { $X_1$};
  \node (X2) at (1.2,-0.85) { $X_2$};
 \end{tikzpicture}
    \caption{Sunrise with two massive and one massless propagators.}
    \label{fig:sunrise}
\end{figure}

Two-mass sunrise integral family in $D=2-2\epsilon$ dimension is defined as
\begin{align}\label{sunrise}
    I_{k_1,k_2,k_3,k_4,k_5}&=\int\frac{{\rm d}^Dy{\rm d}^D z}{(i\pi^{D/2})^2}\frac{((z{-}x_1)^2{-}m_1^2)^{-k_4}((y-x_2)^2-m_2^2)^{-k_5}}{((y{-}x_1)^{2}{-}m_1^2)^{k_1}((y-z)^2)^{k_2}((z{-}x_2)^2{-}m_2^2)^{k_3}}\\
    &=\int\frac{{\rm d}^{D+2}Y{\rm d}^{D+2}Z}{(i\pi^{D/2})^2(GL(1))^2}\frac{(Y,I_\infty)^{-\rho_1}(Z,I_\infty)^{-\rho_2}(Z,X_1)^{-k_4}(Y,X_2)^{-k_5}}{(Y,X_1)^{k_1}(Y,Z)^{k_2}(Z,X_2)^{k_3}}
\end{align}
with $\rho_1=-k_1-k_2-k_5-2\epsilon$, $\rho_2=-k_3-k_2-k_4-2\epsilon$ (see in Fig.~\ref{fig:sunrise}), where we have $p^2=s$, and masses of the two propagators except for the middle massless one, $m_1^2$, $m_2^2$. The integration in embedding space is performed over the light cone $(Y,Y)=(Z,Z)=0$. Once propagator $(y-z)^2$ becomes massive, the integral turns out to give elliptic MPL functions~\cite{Bogner:2019lfa,Wilhelm:2022wow}, which we will comment in section~\ref{discussion}, and here the integral still gives MPL function. To build a complete basis for invariants formed by $\{y,z,x_i\}_{i=1,2}$, in addition to $D_1=(y{-}x_1)^{2}{-}m_1^2$, $D_2=(y-z)^2$ and $D_3=(z{-}x_2)^2{-}m_2^2$, we include two ISPs $D_4=(z{-}x_1)^2{-}m_1^2$ and $D_5=(y-x_2)^2-m_2^2$. Similar to the propagators, they are nicely rewritten in embedding variables as $(Z,X_1)/(Z,I_\infty)$ and $(Y,X_2)/(Y,I_\infty)$. Explicit computation shows that there are {\it four} independent master integrals in this sector. On the other hand, following the topology of the integral we have different independent Schubert problems as well. We will see an elegant one-to-one correspondence between master integrals and Schubert problems!

\paragraph{Schubert problems for all four master integrals}
Let us begin with the integral without ISP, {\it i.e.}
\begin{align}
    I_{1,1,1,0,0}=\int\frac{{\rm d}^{D+2}Y{\rm d}^{D+2}Z}{(i\pi^{D/2})^2(GL(1))^2}\frac{(Y,I_\infty)^{2\epsilon}(Z,I_\infty)^{2\epsilon}}{(Y,X_1)(Y,Z)(Z,X_2)}
\end{align}
We define its Schubert problem in $d=2$ by sending all propagators on shell. Naively, the diagram does not have $2L$ propagators to uniquely define the maximal cut, but it is well known that one can take more residues using the Jacobian from previous steps. Explicitly, when solving conditions $(Y,X_1)=(Y,Z)=0$ at $d=2$, we are actually dealing with a bubble Schubert problem for $Y$. Leading singularity after solving the conditions for the bubble ($2$-gon) in $d=2$ reads 
\[LS\propto1/(\det G(Z,X_1))^\frac12\propto 1/(Z,X_1),\]
which contributes one more condition so that we can continually send $(Z,X_1)=0$ to finally fix the loop momenta. Therefore, Schubert problem of the diagram reads
\begin{equation}\label{prob1}
    (Y,X_1)=(Y,Z)=(Z,X_2)=(Z,X_1)=0\,.
\end{equation}
With a little computation we can figure out that the solutions of $Y$ and $Z$ are actually $Y=Z=Y^2_+$ or $Y=Z=Y^2_-$, where $(Y^2_\pm,X_1)=(Y^2_\pm,X_2)=0$ are just the solutions coinciding with one-loop $d=2$ bubble. Finally, LS for the integral also coincides with the one-loop bubble, {\it i.e.} $1/\sqrt{\mathcal{G}_0^0}=1/\sqrt{\lambda(s,m_1^2,m_2^2)}$ where $\lambda(x,y,z)$ is the K\"{a}ll\'{e}n function, $\lambda(x,y,z)=x^2+y^2+z^2-2xy-2yz-2xz$, and
\begin{equation}
    I_2{=}\sqrt{\lambda(m_1^2,m_2^2,s)}I_{1,1,1,0,0}
\end{equation}
is one of the UT basis. It is well known that loop integrals can develop new poles on the support of other residues, which are called to have {\it composite} leading singularities. Readers can find more examples of this type in the previous investigation at $d=4$ \cite{He:2022tph}. Therefore, Schubert problem we discussed above is exactly associated to the UT basis $I_2$ here. Superscript for $Y_\pm^2$ here stands for the index of master integral $I_2$.

\begin{figure}[htbp]
    \centering
 \begin{tikzpicture}
 	\draw[black,ultra thick] (-2,0)--(-1,0);
 	\draw[black,ultra thick] (2,0)--(1,0);
 	\filldraw[color=black,fill=white,ultra thick] (0,0) circle (1);
 	\draw[black] (-1,0)--(1,0);
 	\node (Y) at (0,0.5) {$Y$};
 	\node (Z) at (0,-0.5) {$Z$};
 	\node (X1) at (-1.5,1) {$X_1$};
 	\node (X2) at (-1.5,-1) {$X_2$};
 	\draw[red,ultra thick] (0,0.8)--(0,1.2);
 	\draw[red,ultra thick] (0,-0.2)--(0,0.2);
 	
 	\draw[black,line width=2.5pt,->] (2.5,0)--(3,0);
 	
 	\draw[black,ultra thick] (3.5,0)--(4.5,0);
 	\draw[black,ultra thick] (7.5,0)--(6.5,0);
 	\draw[black,ultra thick] (6.5,0) arc (0:-180:1);
 	\draw[blue,ultra thick] (4.5,0)--(6.5,0);
 	\node (Y) at (4,1) {$X_1$};
 	\node (Z) at (5.5,-0.5) {$Z$};
 	\node (X2) at (4,-1) {$X_2$};
 	\draw[red,ultra thick] (5.5,-0.8)--(5.5,-1.2);
 	\draw[red,ultra thick] (5.5,-0.2)--(5.5,0.2);
 	
 	\draw[black,line width=2.5pt,->] (8,0)--(8.5,0);
 	
 	\node (Z) at (10,0.35) {\large $Y=Y_\pm^2$};
 	\node (Z) at (10,-0.35) {\large $Z=Y_\pm^2$};
 \end{tikzpicture}
 \caption{Graphical representation of Schubert problem for $I_2$. Red short lines stand for the on-shell conditions, or cuts; the blue line stands for the Jacobian factor $(Z,X_1)$ after solving conditions for $Y$.}
 \label{probI2}
\end{figure}
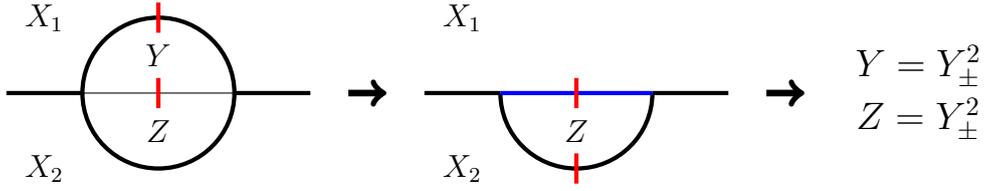

It is worth mentioning that, although the condition $(Z,X_1)$ from the Jacobian apparently breaks the symmetry of the two loop momenta during the procedure,  $Y=Z$ is maintained at the end. In another word, one can alternatively begin with bubble $(Z,X_2)=(Z,Y)=0$ and obtain $(Y,X_2)=0$ as the extra condition, but $(Z,X_1)=0$ is also potentially implied in the problem, leading to the same Schubert solutions. 

Now let us turn to other possible Schubert problems from general $I_{k_1,k_2,k_3,k_4,k_5}$. As we are familiar from one-loop case, besides on-shell conditions from propagators, one can also take $(Y,I_\infty)=0$. Similarly we construct Schubert problems of the form
\begin{equation}\label{prob2}
(Y,X_1)=(Y,Z)=(Z,X_2)=(Z,I_\infty)=0
\end{equation}
With the extra condition $(Z,I_\infty)=0$, we fix all degrees of freedom for loop momenta. Such conditions account for the maximal cut of a master integral with one more ISP in the numerator, which leads to one more $(Z,I_\infty)$ factor in the denominator. For instance, with $\{k_1,k_2,k_3,k_4,k_5\}=\{1,1,1,-1,0\}$, the integral reads
\begin{align}
    I_3=I_{1,1,1,-1,0}=\int\frac{{\rm d}^{D+2}Y{\rm d}^{D+2}Z}{(i\pi^{D/2})^2(GL(1))^2}\frac{(Y,I_\infty)^{2\epsilon}(Z,I_\infty)^{2\epsilon}(Z,X_1)}{(Y,X_1)(Y,Z)(Z,X_2)(Z,I_\infty)}
\end{align}
whose maximal-cut conditions exactly correspond to the Schubert problem \eqref{prob2}. From our one-loop discussion, conditions $(Z,X_2)=(Z,I_\infty)=0$ coincide with those for one-loop tadpole near $d=2$, and we denote its solutions as $Z^3_\pm$. Plugging the solutions for $Z$ into conditions for $Y$, we get two pairs of solutions, which will be denoted as $Y^{3,\pm}_\pm$, satisfying $(Y^{3,+}_\pm,Z^3_+)=(Y^{3,-}_\pm,Z^3_-)=0$ in the following. Similarly, by symmetry we have an alternative Schubert problem as 
\begin{equation}\label{prob3}
(Y,X_1)=(Y,Z)=(Z,X_2)=(Y,I_\infty)=0
\end{equation}
which comes from the maximal cut of
\begin{align}
    I_4=I_{1,1,1,0,-1}=\int\frac{{\rm d}^{D+2}Y{\rm d}^{D+2}Z}{(i\pi^{D/2})^2(GL(1))^2}\frac{(Y,I_\infty)^{2\epsilon}(Z,I_\infty)^{2\epsilon}(Y,X_2)}{(Y,X_1)(Y,Z)(Z,X_2)(Y,I_\infty)}
\end{align}
Solutions from its Schubert problem are denoted as $Y^4_\pm$, $Z^{4,\pm}_\pm$, following the similar procedure. These two integrals $I_3$ and $I_4$ are both UT basis in the top sector. 

Finally, considering $(Y_i,I_\infty)=0$ for $i=1,2$ at the same time, non-trivial Schubert problem can be
\begin{equation}\label{prob4}
     (Y,X_1)=(Y,I_\infty)=(Z,X_2)=(Z,I_\infty)=0
\end{equation}
Obviously, these on-shell condition can be interpreted as maximal cuts of a double-tadpole integral as
\begin{equation}
    I_1=I_{1,0,1,0,0}=\int\frac{{\rm d}^{D+2}Y{\rm d}^{D+2}Z}{(i\pi^{D/2})^2(GL(1))^2}\frac{(Y,I_\infty)^{2\epsilon}}{(Y,X_1)(Y,I_\infty)}\frac{(Z,I_\infty)^{2\epsilon}}{(Z,X_2)(Z,I_\infty)}
\end{equation}
which is just the last UT integral in the sub-sector. Solutions from \eqref{prob4} are again those for one-loop tadpoles, which are $Z_\pm^3$ and $Y_\pm^4$ above.

\begin{figure}[htbp]
    \centering
 \begin{tikzpicture}
 	\draw[black,ultra thick] (-1.25,0)--(0,0);
 	\draw[black,ultra thick] (1.25,0)--(0,0);
    \draw (0,0) [ultra thick]to[out=150,in=180] (0,1);
 	\draw (0,0) [ultra thick]to[out=30,in=0] (0,1);
    \draw (0,0) [ultra thick]to[out=210,in=180] (0,-1);
 	\draw (0,0) [ultra thick]to[out=-30,in=0] (0,-1);
    \draw[black] (-1,0)--(1,0);
 	\node (Y) at (0,0.5) {$Y$};
 	\node (Z) at (0,-0.5) {$Z$};
 	\node (X1) at (-1.5,1) {$X_1$};
 	\node (X2) at (-1.5,-1) {$X_2$};
    \node (X1) at (1.5,1) {$(I_\infty)$};
 	\node (X2) at (1.5,-1) {$(I_\infty)$};
 	\draw[red,ultra thick] (0,0.8)--(0,1.2);
 	\draw[red,ultra thick] (0,-0.8)--(0,-1.2);
    \draw[red,ultra thick] (1.1,1)--(1.9,1);
 	\draw[red,ultra thick] (1.1,-1)--(1.9,-1);
    \node (Z) at (0,-2) {$I_1$};
 	 	
 	\draw[black,ultra thick] (3.5,0)--(4.5,0);
 	\draw[black,ultra thick] (7.5,0)--(6.5,0);
 	\filldraw[color=black,fill=white,ultra thick] (5.5,0) circle (1);
 	\draw[black] (4.5,0)--(6.5,0);
 	\node (Y) at (5.5,0.5) {$Y$};
 	\node (Z) at (5.5,-0.5) {$Z$};
 	\node (X1) at (4,1) {$X_1$};
 	\node (X2) at (4,-1) {$X_2$};
    \node (X1) at (7,1) {$(I_\infty)$};
 	\node (X2) at (7,-1) {$(I_\infty)$};
 	\draw[red,ultra thick] (5.5,0.8)--(5.5,1.2);
 	\draw[red,ultra thick] (5.5,-0.8)--(5.5,-1.2);
    \draw[red,ultra thick] (5.5,-0.2)--(5.5,0.2);
 	\draw[red,ultra thick] (6.6,-1)--(7.4,-1);
    \node (Z) at (5.5,-2) {$I_3$};
 	
 	\draw[black,ultra thick] (9,0)--(10,0);
 	\draw[black,ultra thick] (13,0)--(12,0);
 	\filldraw[color=black,fill=white,ultra thick] (11,0) circle (1);
 	\draw[black] (10,0)--(12,0);
 	\node (Y) at (11,0.5) {$Y$};
 	\node (Z) at (11,-0.5) {$Z$};
 	\node (X1) at (9.5,1) {$X_1$};
 	\node (X2) at (9.5,-1) {$X_2$};
    \node (X1) at (12.5,1) {$(I_\infty)$};
 	\node (X2) at (12.5,-1) {$(I_\infty)$};
 	\draw[red,ultra thick] (11,0.8)--(11,1.2);
 	\draw[red,ultra thick] (11,-0.8)--(11,-1.2);
    \draw[red,ultra thick] (11,-0.2)--(11,0.2);
 	\draw[red,ultra thick] (12.1,1)--(12.9,1);
    \node (Z) at (11,-2) {$I_4$};
 \end{tikzpicture}
 \caption{Schubert problems for $I_1$, $I_3$ and $I_4$. Slashed $I_\infty$ on the upper (lower) side stands for condition $(Y,I_\infty)=0$ ($(Z,I_\infty)=0$).}
 \label{probI134}
\end{figure}
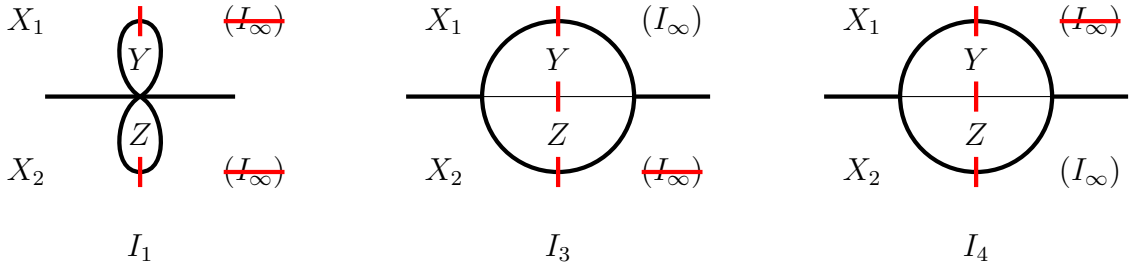

In summary, we discover an exact one-to-one correspondence between $4$ Schubert problems associated with the topology and four independent UT basis in the sector (see fig.\ref{probI2} and fig.\ref{probI134}). In the next part we generate all possible symbol letters from these Schubert solutions, and we will see that they reproduce those needed in the differential equations. 

It is also worth mentioning that all these can be carried out under Baikov representation of the $4$ master integrals as well. To be a bit more explicit, in Baikov representation, each of the four UT integrals above can be expressed as a single ${\rm d}\log$ form as
\[I=\mathcal{N}\int\prod_{i=1}^4{\rm d}\log(f_i)\]
with  $f_1,\cdots,f_4$ being functions of $D_i$. Maximal cut conditions under this representation are simply $f_1=\cdots =f_4=0$, which exactly correspond to the on-shell conditions of each Schubert problem. We record more details about this procedure in Appendix \ref{Baikov}.


\paragraph{Symbol letters and differential equations}
Now we reproduce symbol letters and compare with canonical differential equations. As mentioned above, a simple computation from integration by part shows that there are exactly four master integrals in this family, which can be chosen as the following four UT basis:
\begin{equation}
    \begin{aligned}
        I_{1}&=I_{1,0,1,0,0}, \ I_{2}=\sqrt{\lambda(s,m_{1}^{2},m_{2}^{2})}I_{1,1,1,0,0}, \\
        I_{3}&=I_{1,1,1,-1,0}, \ I_{4}=I_{1,1,1,0,-1}.
    \end{aligned}
\end{equation}
Canonical differential equation (CDE) for this basis can be computed as
\begin{equation}\label{CDE2}
    \mathrm{d}\mathbf{I}=\epsilon \mathrm{d}A \mathbf{I}, \, 
    A=\left(\begin{array}{cccc}
        -R_{1}-R_{2} & 0 & 0 & 0 \\
         -R_{5} & R_{1}+R_{2}+2R_{3}-3R_{4} & R_{5}-R_{7} & R_{5}-R_{6} \\
         -R_{1}+R_{2} & R_{7} & -2R_{2} & R_{3}-R_{2} \\
         R_{1}-R_{2} & R_{6} & R_{3}-R_{1} & -2R_{1}
    \end{array}\right) \, .
\end{equation}
where $R_{i}=\log W_{i}$, and the letters $W_i$ are defined as
\begin{equation}
    \begin{aligned}
        W_{1}&=m_1^2, \, W_{2}=m_{2}^2, \, W_{3}=s, \, W_{4}=\lambda(s,m_1^2,m_2^2) \\
        W_{5}&=\frac{m_1^2+m_2^2-s-\sqrt{\lambda(s,m_1^2,m_2^2)}}{m_1^2+m_2^2-s+\sqrt{\lambda(s,m_1^2,m_2^2)}} , \\
        W_{6}&=\frac{s+m_1^2-m_2^2-\sqrt{\lambda(s,m_1^2,m_2^2)}}{s+m_1^2-m_2^2+\sqrt{\lambda(s,m_1^2,m_2^2)}}, \\
        W_{7}&=\frac{s+m_2^2-m_1^2-\sqrt{\lambda(s,m_1^2,m_2^2)}}{s+m_2^2-m_1^2+\sqrt{\lambda(s,m_1^2,m_2^2)}}\,.
    \end{aligned}
\end{equation}
Note that $W_{5},W_{6},W_{7}$ are not independent, actually
\begin{equation}
    R_{5}+R_{6}+R_{7}=0 \, .
\end{equation}
We keep this redundancy to put the CDE into a nicer form.

It is straightforward to reproduce these letters from Schubert analysis. Two classes of letters arise naturally from the analysis; (1): letters by taking product of two solutions from one individual problem, and (2) letters by combining solutions of two distinct Schubert problems. The first class accounts for the four even letters and the diagonal elements in \eqref{CDE2}. Note that  we still have the ambiguity up to the fixing of projectivity. For instance, we choose to parametrize each solutions always by $X_1+a_2 X_2+a_p Y^2_++a_m Y^2_-$ and construct all the inner products, then
\begin{align}
    &(Y^2_+,Y^2_-)=\frac{W_4}{W_3},\ \  (Z^3_+,Z^3_-)=\frac{W_4^2}{W_3^2W_1},\ \  (Y^4_+,Y^4_-)=\frac{W_4^2}{W_3^2W_2}\\
    &(Z^{4,+}_+,Z^{4,+}_-)(Z^{4,-}_+,Z^{4,-}_-){=}(Y^{3,+}_+,Y^{3,+}_-)(Y^{3,-}_+,Y^{3,-}_-)= \frac{W_4^4}{W_3^2W_1^2W_2^2}
\end{align}
which give all four non-trivial even letters. On the other hand, letters from the second class contain odd letters $W_5,W_6,W_7$. For instance, combining Schubert solutions from $I_2$ and $I_3$ like \eqref{type1}, we have the following odd letters
\begin{align}
 &\frac{(Y^2_+,Z^3_+)(Y^2_-,Z^3_-)}{(Y^2_+,Z^3_-)(Y^2_-,Z^3_+)}{=}W_7^2,\, \frac{(Y^2_+,Y^{3,+}_+)(Y^2_-,Y^{3,-}_+)}{(Y^2_+,Y^{3,-}_+)(Y^2_-,Y^{3,+}_+)}{=}\frac{W_5^2}{W_7^2}\label{W7}\\
    &\frac{(Y^2_+,Y^{3,-}_-)(Y^2_-,Y^{3,+}_-)}{(Y^2_+,Y^{3,+}_-)(Y^2_-,Y^{3,-}_-)}{=}{W_5^2}{W_7^2},\,\frac{(Y^2_+,Y^{3,+}_+)(Y^2_-,Y^{3,+}_-)}{(Y^2_+,Y^{3,+}_-)(Y^2_-,Y^{3,+}_+)}{=}\frac{(Y^2_+,Y^{3,-}_-)(Y^2_-,Y^{3,-}_+)}{(Y^2_+,Y^{3,-}_+)(Y^2_-,Y^{3,-}_-)}{=}W_5^2
\end{align}
We see that \eqref{W7} exactly reproduce the dependence of ${\rm d}I_2$ on $I_3$ and ${\rm d}I_3$ on $I_2$. As for combining $I_2$ and $I_4$, we get $W_6$ and $W_6/W_5$ instead. We can also consider possible cross-ratios from combining $\{Z^3_\pm,Y^{3,\pm}_\pm\}$, $\{Y^4_\pm,Z^{4,\pm}_\pm\}$. Most of those read $0$ or $\infty$, while cross-ratios similar to the second in \eqref{W7} read
\begin{equation}
    \frac{(Y^4_+,Y^{3,+}_+)(Y^4_-,Y^{3,-}_+)}{(Y^4_+,Y^{3,-}_+)(Y^4_-,Y^{3,+}_+)}=\frac{W_1^2}{W_3^2},\quad \frac{(Z^3_+,Z^{4,+}_+)(Z^3_-,Z^{4,-}_+)}{(Z^3_+,Z^{4,-}_+)(Z^3_-,Z^{4,+}_+)}=\frac{W_2^2}{W_3^2}
\end{equation}
exactly reproducing the dependence of ${\rm d}I_3$ on $I_4$ and ${\rm d}I_4$ on $I_3$! This is the first evidence that Schubert analysis can not only recover the alphabet, but also see where those letters appear in the CDE.

\subsection{Four-mass double-box and double-triangle integrals in $D=4-2\epsilon$}
Next we move to more non-trivial examples in the family of double-box integrals in $D=4-2\epsilon$ with four massive external legs, and their canonical differential equations have been determined in~\cite{He:2022ctv}. We will focus on two sectors, {\it i.e.}, the top sector for double-box integrals containing $4$ master integrals, and the double-triangle sector containing $7$ master integrals. Note that in principle all symbol letters of the family were derived using Schubert analysis in twistor space in~\cite{He:2022tph}, but here we perform Schubert analysis in embedding space and find precisely how these letters appear in CDE. An important difference from one-loop and the sunrise cases is that it is no longer the case that each UT master integral is associated with an individual Schubert problem. Nevertheless, we will see that we can still interpret all symbol letters from Schubert analysis in embedding space. 

\begin{figure}[htbp]
\centering
\includegraphics[width=0.5\textwidth]{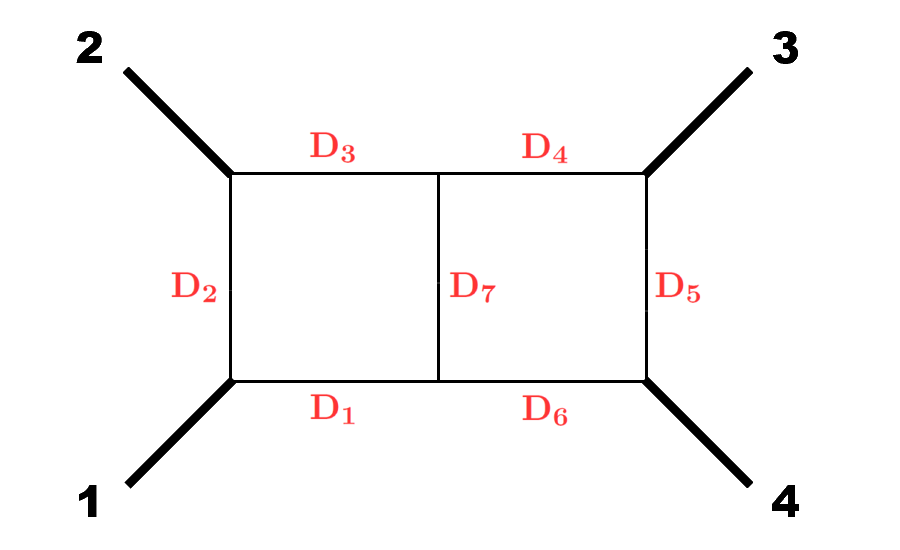}
\caption{double-box integral with 4 external massive legs}
\label{Figure_dbox}
\end{figure}

Let us first begin with basic definition of the integral family. Following~\cite{He:2022ctv}, the double-box integral family is defined by nine ISPs as
\begin{align}
    D_1=(y-x_1)^2,\ D_2=(y-x_2)^2,\ D_3=(y-x_3)^2 \nonumber\\
    D_4=(z-x_3)^2,\ D_5=(z-x_4)^2,\ D_6=(z-x_1)^2 \nonumber\\
    D_7=(y-z)^2,\ D_8=(y-x_4)^2,\ D_9=(z-x_2)^2 
\end{align}
and the integral family reads
\begin{equation}\label{doublebox}
    J_{a_1,\cdots,a_9}=\int\frac{{\rm d}^Dy{\rm d}^Dz}{(i\pi^{D/2})^2}\frac{D_8^{-a_8}D_9^{-a_9}}{D_1^{a_1}D_2^{a_2}D_3^{a_3}D_4^{a_4}D_5^{a_5}D_6^{a_6} D_7^{a_7}}
\end{equation}
Explicit computation from differential equations shows that there are $74$ master integrals and $68$ multiplicatively independent MPL symbol letters in the differential equation system. It is also straightforward to rewrite $J[a_1,\cdots,a_9]$ by embedding variables as
\begin{equation}
   \int\frac{{\rm d}^{D+2}Y{\rm d}^{D+2}Z}{(i\pi^{D/2})^2(GL(1))^2}\frac{(Y,I_\infty)^{-\rho_1}(Z,I_\infty)^{-\rho_2}(Y,X_4)^{-a_8}(Z,X_2)^{-a_9}}{(Y,X_1)^{a_1}(Y,X_2)^{a_2}(Y,X_3)^{a_3}(Z,X_3)^{a_4}(Z,X_4)^{a_5}(Z,X_1)^{a_6} (Y,Z)^{a_7}}
\end{equation}
with $\rho_1=-a_1-a_2-a_3-a_7-a_8-2\epsilon$, $\rho_2=-a_4-a_5-a_6-a_7-a_9-2\epsilon$, and the integration is again under the conditions $(Y,Y)=(Z,Z)=0$.

\paragraph{Double-box integrals}
Let us first study the top sector $J_{1,1,1,1,1,1,1,0,0}$, which contains four UT master integrals which we denote as
\begin{align}
    &J_1=sr_1J_{1,1,1,1,1,1,1,0,0}\nonumber\\
    &J_2=r_4({-}m_1^2J_{1,1,1,0,1,1,1,0,0}{-}m_2^2J_{1,1,1,1,1,0,1,0,0}{+}sJ_{1,1,1,1,1,1,1,0,-1})\nonumber\\
    &J_3=r_2({-}m_3^2J_{0,1,1,1,1,1,1,0,0}{-}m_4^2J_{1,1,0,1,1,1,1,0,0}{+}sJ_{1,1,1,1,1,1,1,-1,0})\nonumber \\
    &J_4{=}sJ_{1,1,1,1,1,1,1,-1,-1}{+}\frac12s(s{-}m_1^2{-}m_2^2)J_{1,1,1,1,1,1,1,-1,0}{+}\frac12t(t{-}m_3^2{-}m_4^2)J_{1,1,1,1,1,1,1,0,-1}{+}\cdots
\end{align}
with $r_i$ being leading singularities of the one-loop box ($r_1$) and four triangles $r_2\sim r_5$ in \eqref{defi:r1-5}, and the omission for $J_4$ denotes subsector integrals. We refer to appendix \ref{4mletters} for the definition of all square roots and letters. 

Now we consider applying our analysis to this sector. In our previous examples, we see that each integral has unique maximal cut, which leads to a unique Schubert problem. This is the case for $J_1$, $J_2$ and $J_3$, which has less than $8$ propagators. We find that their maximal cuts correspond to cutting all propagators, and $(Y,I_\infty)$ or $(Z,I_\infty)$ which appear due to the ISP factors of $J_2$ and $J_3$:
\begin{align}\label{eq:dbprob}
    J_1&: (Y,X_1)=(Y,X_2)=(Y,X_3)=(Y,Z)=(Z,X_3)=(Z,X_4)=(Z,X_1)=0\nonumber\\
    J_2&: (Y,X_1)=(Y,X_2)=(Y,X_3)=(Y,Z)=(Z,X_3)=(Z,X_4)=(Z,X_1)=(Z,I_\infty)=0\nonumber\\
     J_3&: (Y,X_1)=(Y,X_2)=(Y,X_3)=(Y,Z)=(Z,X_3)=(Z,X_4)=(Z,X_1)=(Y,I_\infty)=0
\end{align}

\begin{figure}[htbp]
    \centering
 \begin{tikzpicture}[scale=0.65]
 	\draw[black] (0,-1)--(-2,-1)--(-2,1)--(0,1)--(0,-1)--(2,-1)--(2,1)--(0,1)--cycle;
 	\draw[black,ultra thick] (-2,1)--(-3,2);
 	\draw[black,ultra thick] (-2,-1)--(-3,-2);
    \draw[black,ultra thick] (2,1)--(3,2);
    \draw[black,ultra thick] (2,-1)--(3,-2);
    \node (X1) at (-1,0) {$Y$};
    \node (X1) at (1,0) {$Z$};
    \node (X1) at (0,-1.75) {$X_1$};
    \node (X1) at (0,1.75) {$X_3$};
    \node (X1) at (-2.75,0) {$X_2$};
    \node (X1) at (2.75,0) {$X_4$};
    \node (X1) at (-1.5,-2) {$(I_{\infty})$};
    \node (X1) at (1.5,-2) {$(I_{\infty})$};
    \node (X1) at (0,-3.25) {$J_1$};
    \draw[red,ultra thick] (-2.2,0)--(-1.8,0);
    \draw[red,ultra thick] (1.8,0)--(2.2,0);
    \draw[red,ultra thick] (-1,0.8)--(-1,1.2);
    \draw[red,ultra thick] (-1,-0.8)--(-1,-1.2);
    \draw[red,ultra thick] (1,-0.8)--(1,-1.2);
    \draw[red,ultra thick] (1,0.8)--(1,1.2);
    \draw[red,ultra thick] (-0.2,0.1)--(0.2,0.1);
    \draw[red,ultra thick] (-0.2,-0.1)--(0.2,-0.1);

    \draw[black] (7.5,-1)--(5.5,-1)--(5.5,1)--(7.5,1)--(7.5,-1)--(9.5,-1)--(9.5,1)--(7.5,1)--cycle;
 	\draw[black,ultra thick] (5.5,1)--(4.5,2);
 	\draw[black,ultra thick] (5.5,-1)--(4.5,-2);
    \draw[black,ultra thick] (9.5,1)--(10.5,2);
    \draw[black,ultra thick] (9.5,-1)--(10.5,-2);
    \node (X1) at (6.5,0) {$Y$};
    \node (X1) at (8.5,0) {$Z$};
    \node (X1) at (7.5,-1.75) {$X_1$};
    \node (X1) at (7.5,1.75) {$X_3$};
    \node (X1) at (4.75,0) {$X_2$};
    \node (X1) at (10.25,0) {$X_4$};
    \node (X1) at (6,-2) {$(I_{\infty})$};
    \node (X1) at (9,-2) {$(I_{\infty})$};
    \node (X1) at (7.5,-3.25) {$J_2$};
    \draw[red,ultra thick] (-2.2+7.5,0)--(-1.8+7.5,0);
    \draw[red,ultra thick] (1.8+7.5,0)--(2.2+7.5,0);
    \draw[red,ultra thick] (-1+7.5,0.8)--(-1+7.5,1.2);
    \draw[red,ultra thick] (-1+7.5,-0.8)--(-1+7.5,-1.2);
    \draw[red,ultra thick] (1+7.5,-0.8)--(1+7.5,-1.2);
    \draw[red,ultra thick] (1+7.5,0.8)--(1+7.5,1.2);
    \draw[red,ultra thick] (-0.2+7.5,0)--(0.2+7.5,0);
    \draw[red,ultra thick] (8.5,-2)--(9.5,-2);

    \draw[black] (15,-1)--(13,-1)--(13,1)--(15,1)--(15,-1)--(17,-1)--(17,1)--(15,1)--cycle;
 	\draw[black,ultra thick] (13,1)--(12,2);
 	\draw[black,ultra thick] (13,-1)--(12,-2);
    \draw[black,ultra thick] (17,1)--(18,2);
    \draw[black,ultra thick] (17,-1)--(18,-2);
    \node (X1) at (14,0) {$Y$};
    \node (X1) at (16,0) {$Z$};
    \node (X1) at (15,-1.75) {$X_1$};
    \node (X1) at (15,1.75) {$X_3$};
    \node (X1) at (12.25,0) {$X_2$};
    \node (X1) at (17.75,0) {$X_4$};
    \node (X1) at (13.5,-2) {$(I_{\infty})$};
    \node (X1) at (16.5,-2) {$(I_{\infty})$};
    \node (X1) at (15,-3.25) {$J_3$};
    \draw[red,ultra thick] (-2.2+15,0)--(-1.8+15,0);
    \draw[red,ultra thick] (1.8+15,0)--(2.2+15,0);
    \draw[red,ultra thick] (-1+15,0.8)--(-1+15,1.2);
    \draw[red,ultra thick] (-1+15,-0.8)--(-1+15,-1.2);
    \draw[red,ultra thick] (1+15,-0.8)--(1+15,-1.2);
    \draw[red,ultra thick] (1+15,0.8)--(1+15,1.2);
    \draw[red,ultra thick] (-0.2+15,0)--(0.2+15,0);
    \draw[red,ultra thick] (-0.2,-0.1)--(0.2,-0.1);
    \draw[red,ultra thick] (13,-2)--(14,-2);
 \end{tikzpicture}
 \caption{Schubert problems for $J_1$, $J_2$ and $J_3$. Double-slashed propagators stands for composite-type leading singularity and in this special case we have $(Z,X_2)=0$ or $(Y,X_4)=0$.}
\end{figure}
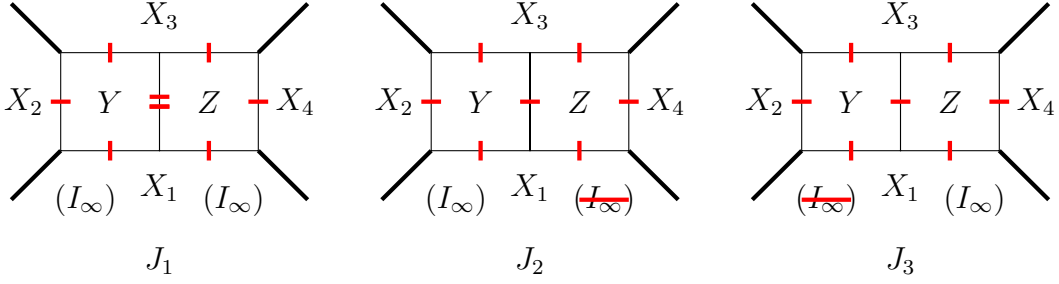
However, $J_4$ is different: it has $9$ factors $D_1\cdots D_7$ as well as $(Y,I_\infty)$ and $(Z,I_\infty)$ on its denominator following from  $J_{1,1,1,1,1,1,1,-1,-1}$. Number of these factors is more than the degrees of freedom of $Y$ and $Z$. Therefore $J_4$ does not have a unique maximal cut, so that we cannot associate a unique Schubert problem to it. 

Let us look at these Schubert problems more carefully. The leading singularity of $J_1$ is still composite just like $I_2$ in sunrise case: the loop momenta are fully determined with the conditions from Jacobian factor, either $(Z,X_2)=0$ or $(Y,X_4)=0$ from solving on-shell conditions for $Y$ or $Z$ first, respectively; $Y$ and $Z$ are both localized at solutions for one-loop four-mass boxes, which are denoted as $Y^1_\pm$ in this section. For $J_2$ ($J_3$), we first have triangle on-shell conditions for $Z$ ($Y$), which give two solutions $Z^2_\pm$ ($Y^3_\pm$); by plugging the solution into $(Y,Z)$ and solving the rest one, we get $Y^{2,\pm}_\pm$ ($Z^{3,\pm}_\pm$), which are similar to $I_3$ and $I_4$ in the sunrise case. That is why $J_1$ has one-loop box leading singularity, and $J_2$ and $J_3$ have one-loop triangle leading singularities. 

Remarkably, solutions of each individual Schubert problem exactly contribute to the corresponding ${\rm d}\log$ coefficients in the differential equation! For instance, in the top sector we have the relation
\begin{align}\label{dJ1}
    {\rm d}J_1=&{\rm d}\log \frac{W_1W_2W_3W_4W_{18}^2}{W_{13}^3} J_1-{\rm d}\log W_{51} J_2-{\rm d}\log W_{46} J_3+2 {\rm d}\log W_{25} J_4\\
    &+\text{lower sector contributions}
\end{align}
Constructing cross-ratios from the solutions, we have (fixing $(Y^1_\pm,I_\infty)=1$)
\begin{align}\label{doubleboxletter}
    &(Y^1_+,Y^1_-)=\frac{W_{13}}{W_{18}},\ \nonumber\\
    &\frac{(Y^1_+,Z^2_+)(Y^1_-,Z^2_-)}{(Y^1_-,Z^2_+)(Y^1_+,Z^2_-)}=W_{51}^2,\ \frac{(Y^1_+,Y^{2,+}_+)(Y^1_-,Y^{2,-}_+)}{(Y^1_-,Y^{2,-}_+)(Y^1_+,Y^{2,+}_+)}=W_{51}^2,\,\frac{(Y^1_+,Y^{2,+}_+)(Y^1_-,Y^{2,+}_-)}{(Y^1_-,Y^{2,+}_+)(Y^1_+,Y^{2,+}_-)}=W_{25}^2
\end{align}
and similar for $W_{46}$ from $J_3$ and its Schubert solutions $\{Y^3_\pm,Z^{3,\pm}_\pm\}$. Comparing with the previous sunrise example, we see that cross-ratios generating $W_{51}$ are exactly the same as \eqref{W7}. The only difference is that, as can be checked from CDE, now dependence of ${\rm d}J_1$ on $J_2$  is the same as ${\rm d}J_2$  on $J_1$, therefore from those cross-ratios we get the same letter $W_{51}$. Finally, although we do not have a unique Schubert problem for $J_4$, its coefficient in \eqref{dJ1} , (${\rm d} \log$ of) $W_{25}$, has already been produced in \eqref{doubleboxletter}. 

\begin{figure}[htbp]
    \centering
 \begin{tikzpicture}[scale=0.65]
 	\draw[black] (0,-1)--(-2,-1)--(-2,1)--(0,1)--(0,-1)--(2,0)--(0,1)--cycle;
 	\draw[black,ultra thick] (-2,1)--(-3,2);
 	\draw[black,ultra thick] (-2,-1)--(-3,-2);
    \draw[black,ultra thick] (2,0)--(3,0);
    \draw[black,ultra thick] (0,-1)--(0.75,-2);
    \node (X1) at (-1,0) {$Y$};
    \node (X1) at (0.75,0) {$Z$};
    \node (X1) at (-2.75,0) {$X_2$};
    \node (X1) at (0,1.5) {$X_3$};
    \node (X1) at (-1,-1.75) {$X_1$};
    \node (X1) at (0.75+1,-1) {$X_4$};
    \node (X1) at (-1-0.75,-2-0.5) {$(I_{\infty})$};
    \node (X1) at (2,-2.5) {$(I_{\infty})$};
    \node (X1) at (0,-3.25) { $J_8$};
    \draw[red,ultra thick] (-2.2,0)--(-1.8,0);
    \draw[red,ultra thick] (-1,0.8)--(-1,1.2);
    \draw[red,ultra thick] (-1,-0.8)--(-1,-1.2);
    \draw[red,ultra thick] (-0.2,0.1)--(0.2,0.1);
    \draw[red,ultra thick] (-0.2,-0.1)--(0.2,-0.1);
    \draw[red,ultra thick] (1-0.2,0.5-0.2)--(1+0.2,0.5+0.2);
    \draw[red,ultra thick] (1-0.2,-0.5+0.2)--(1+0.2,-0.5-0.2);
    \draw[red,ultra thick] (1.5,-2.5)--(2.5,-2.5);

    \draw[black] (0+8,-1)--(-2+8,-1)--(-2+8,1)--(0+8,1)--(0+8,-1)--(2+8,0)--(0+8,1)--cycle;
 	\draw[black,ultra thick] (-2+8,1)--(-3+8,2);
 	\draw[black,ultra thick] (-2+8,-1)--(-3+8,-2);
    \draw[black,ultra thick] (2+8,0)--(3+8,0);
    \draw[black,ultra thick] (0+8,-1)--(0.75+8,-2);
    \node (X1) at (-1+8,0) {$Y$};
    \node (X1) at (0.75+8,0) {$Z$};
    \node (X1) at (-2.75+8,0) {$X_2$};
    \node (X1) at (0+8,1.5) {$X_3$};
    \node (X1) at (-1+8,-1.75) {$X_1$};
    \node (X1) at (0.75+1+8,-1) {$X_4$};
    \node (X1) at (-1-0.75+8,-2-0.5) {$(I_{\infty})$};
    \node (X1) at (2+8,-2.5) {$(I_{\infty})$};
    \node (X1) at (3,0.3) {$m_3^2$};
    \node (X1) at (0+8,-3.25) {$J_{29}$};
    \draw[red,ultra thick] (-2.2+8,0)--(-1.8+8,0);
    \draw[red,ultra thick] (-1+8,0.8)--(-1+8,1.2);
    \draw[red,ultra thick] (-1+8,-0.8)--(-1+8,-1.2);
    \draw[red,ultra thick] (-0.2+8,0)--(0.2+8,0);
    \draw[red,ultra thick] (1-0.2+8,0.5-0.2)--(1+0.2+8,0.5+0.2);
    \draw[red,ultra thick] (1-0.2+8,-0.5+0.2)--(1+0.2+8,-0.5-0.2);
    \draw[red,ultra thick] (1.5+8,-2.5)--(2.5+8,-2.5);
    \draw[red,ultra thick] (-1-0.75+8-0.5,-2-0.5)--(-1-0.75+8+0.5,-2-0.5);
 \end{tikzpicture}
 \caption{Schubert problems for $J_8$ and $J_{29}$. After $Z$ is determined in the two cases, besides the three on-shell conditions from propagators for $Y$, in $J_8$ we solve jacobian condition $(Y,X_4)-m_3^2=0$, while for $J_{29}$ we consider $(Y,I_\infty)=0$.}
 \label{fig:probboxtri}
\end{figure}
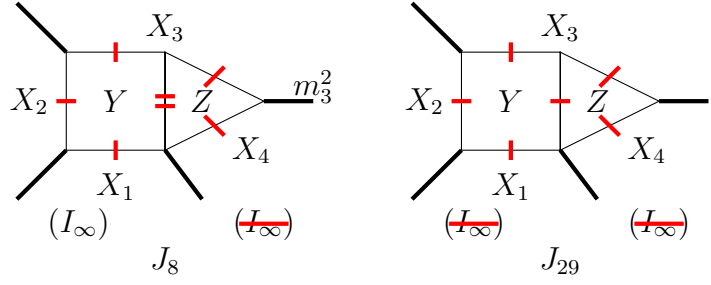

Next we look into contributions from lower sectors in \eqref{dJ1}. There are $26$ extra terms, and these coefficients contain 
\[\{W_1,\cdots,W_4,W_{13},W_{18},W_{25},W_{46},W_{51};W_{65},\cdots,W_{68}\}\]
and the four new letters are from box-triangle sub-sectors (see fig.\ref{fig:probboxtri}). For instance, we have one term
\begin{equation}
    {\rm d}J_1=\cdots-{\rm d}\log W_{67}J_8+\cdots
\end{equation}
and we need another Schubert problem from the box-triangle integral
\begin{align}
     J_8=&\frac{r_{10}J_{1,1,1,1,1,0,1,0,0}}{\epsilon}
\end{align}
by
\begin{equation}\label{prob8}
    (Y,X_1)=(Y,X_2)=(Y,X_3)=(Y,X_4)-m_3^2=(Y,Z)=(Z,X_3)=(Z,X_4)=(Z,I_\infty)=0
\end{equation}
$(Z,I_\infty)=0$ shows up here since we have triangle topology for $Z$. Denoting its solutions by $Z^{8}_\pm$, we have 
\[\frac{(Y^1_+,Z^8_+)(Y^1_-,Z^8_-)}{(Y^1_+,Z^8_-)(Y^1_-,Z^8_+)}=W_{67}^2\]
Other letters $\{W_{65},W_{66},W_{68}\}$ are related to $W_{67}$ by symmetry, which are odd letters containing box-triangle singularities $r_8,r_9,r_{11}$, as recorded in appendix \ref{4mletters}. Therefore, we successfully reproduce all the letters appearing on RHS of ${\rm d}J_1$. 

Similar discussions apply to ${\rm d}J_2$ and ${\rm d}J_3$ as well. For instance, 
\begin{align}
    {\rm d}J_3&={\rm d}\log W_{46}J_1{+}{\rm d}\log W_{35}J_2{-}\frac12{\rm d}\log\frac{W_1W_2W_5^2}{W_9}J_3+2{\rm d}\log W_{21} J_4\nonumber\\
    &+\text{lower sector contributions}
\end{align}
and we have
\begin{align}
    (Y^3_+,Y^3_-)\propto W_9,\, \frac{(Y^3_+,Z^2_-)(Y^3_-,Z^2_+)}{(Y^3_+,Z^2_+)(Y^3_-,Z^2_-)}=W_{35}^2,\, \frac{(Y^3_+,Y^{2,+}_+)(Y^3_-,Y^{2,+}_-)}{(Y^3_-,Y^{2,+}_+)(Y^3_+,Y^{2,+}_-)}=W_{21}^2
\end{align}
More new letters are included in the lower sector contribution part, but they can be similarly recovered by the above procedure.

To close the discussion of this sector,  we consider total differential of $J_4$
\begin{align}
    {\rm d}J_4&=-\frac12{\rm d}\log W_{25} J_1-\frac12{\rm d}\log W_{22} J_2-\frac12{\rm d}\log W_{21}J_3-{\rm d}\log\frac{W_1W_2W_3W_4W_5}{W_{18}}J_4\nonumber\\
    &+\text{lower sector contributions}
\end{align}
Since $J_4$ does not correspond to a single Schubert problem, the above discussion cannot be directly apply to this row of CDE. Nevertheless, all letters for ${\rm d}J_4$ have been reproduced by Schubert analysis for other master integrals. It is an interesting open question to fully understand the structure of CDE for integrals like $J_4$.

\paragraph{Double-triangle integrals}
We move to another interesting sector, namely the double-triangle sector. As computed from IBP relations~\cite{He:2022ctv}, there are two double-triangle sub-sector related by symmetry, and each of them contains $7$ independent UT master integrals, which are (choosing the sector $a_3=a_6=0$ as a representation)
\begin{align}
    &J_{27}{=}\frac{m_1^2r_4J_{2,1,0,1,1,0,1,0,0}}{\epsilon},J_{28}{=}\frac{m_1^2r_5J_{1,2,0,1,1,0,1,0,0}}{\epsilon},J_{29}{=}\frac{m_3^2r_2J_{1,1,0,2,1,0,1,0,0}}{\epsilon}\nonumber\\
    &J_{30}{=}\frac{m_3^2r_3J_{1,1,0,1,2,0,1,0,0}}{\epsilon},J_{31}{=}\frac{r_1J_{1,1,0,1,1,0,2,0,0}}{\epsilon},J_{32}{=}\frac{r_7J_{1,1,0,1,1,0,1,0,0}}{\epsilon}\nonumber\\
    &J_{33}=\frac{s m_1^2m_3^2}{\epsilon^2}J_{2,1,0,2,1,0,1,0,0}+\cdots
\end{align}
The first six of them ($J_{27},\cdots, J_{32}$) have unique maximal cut, and we find the nice correspondence between master integrals and Schubert problems as before, while similar to $J_4$, $J_{33}$ does not have a unique maximal cut. For instance, $J_{32}$ corresponds to the composite leading singularity from
\[(Y,X_1)=(Y,X_2)=(Y,I_\infty)=(Y,Z)=(Z,X_3)=(Z,X_4)=(Z,I_\infty)=0, (Z,X_1)=(Z,X_2)\]
where the first seven conditions are from propagators, while the last one is from Jacobian factor when solving conditions for $Y$ at first. These conditions introduce a novel two-loop leading singularity, which gives $r_7$ above. Explicit computations show that we have a pair of solutions for each loop momentum, denoted as $\{Y^{32}_\pm,Z^{32}_\pm\}$. 

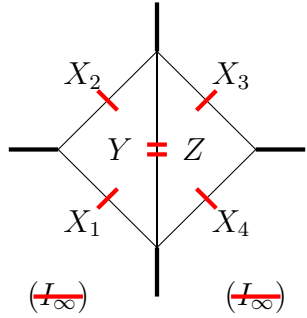
\begin{figure}[htbp]
    \centering
 \begin{tikzpicture}[scale=0.65]
 	\draw[black] (0,2)--(-2,0)--(0,-2)--(0,2)--(2,0)--(0,-2)--cycle;
 	\draw[black,ultra thick] (0,2)--(0,3);
 	\draw[black,ultra thick] (0,-2)--(0,-3);
    \draw[black,ultra thick] (-2,0)--(-3,0);
    \draw[black,ultra thick] (2,0)--(3,0);
    \node (X1) at (-0.75,0) {$Y$};
    \node (X1) at (0.75,0) {$Z$};
    \node (X1) at (-1.5,-1.5) {$X_1$};
    \node (X1) at (-1.5,1.5) {$X_2$};
    \node (X1) at (1.5,1.5) {$X_3$};
    \node (X1) at (1.5,-1.5) {$X_4$};
    \node (X1) at (-2,-3) {$(I_{\infty})$};
    \node (X1) at (2,-3) {$(I_{\infty})$};
    \draw[red,ultra thick] (-0.2,0.1)--(0.2,0.1);
    \draw[red,ultra thick] (-0.2,-0.1)--(0.2,-0.1);
    \draw[red,ultra thick] (1-0.2,1-0.2)--(1+0.2,1+0.2);
    \draw[red,ultra thick] (1-0.2,-1+0.2)--(1+0.2,-1-0.2);
    \draw[red,ultra thick] (-1-0.2,1+0.2)--(-1+0.2,1-0.2);
    \draw[red,ultra thick] (-1+0.2,-1+0.2)--(-1-0.2,-1-0.2);
    \draw[red,ultra thick] (1.5,-3)--(2.5,-3);
    \draw[red,ultra thick] (-1.5,-3)--(-2.5,-3);
 \end{tikzpicture}
 \caption{Schubert problem for $J_{32}$}
\end{figure}
The other five integrals all have double propagators, thus the correspondence with Schubert problems is trickier since naively they are not ${\rm d}\log$ integrals. However, as pointed out by \cite{Dlapa:2021qsl}, a better way to investigate these integrals is converting them by IBP relations to integrals in the supersector which only involve single propagators. For instance, following IBP relations we actually have  
\begin{align}
    J_{29}=\frac{m_3^2r_2J_{1,1,0,2,1,0,1,0,0}}{\epsilon}&=-r_2(J_{1,1,1,1,1,0,1,-1,0}-m_3^2J_{1,1,1,1,1,0,1,0,0}-J_{1,1,0,1,1,0,1,0,0})\nonumber\\
    &+\text{lower sector integrals}
\end{align}
The maximal cut for RHS becomes apparent, which follows from box-triangle topology (fig.\ref{fig:probboxtri}) and gives the Schubert problem~\footnote{It is worth mentioning that there is only one condition $(Y,I_\infty)=0$ in \eqref{prob29} that differs from \eqref{prob8}, which alternatively has $(Y,X_4)-m_3^2=0$.  In fact, $(Y,X_4)-m_3^2$ is the Jacobian factor when we solve four conditions to determine $Z$ in \eqref{prob8}. Therefore \eqref{prob8} accounts for composite leading singularity of $J_8$. On the other hand, $(Y,I_\infty)=0$ in \eqref{prob29} is associated to a box-triangle with ISP on its numerator (see also Fig.\ref{fig:probboxtri}.)} 
\begin{equation}\label{prob29}
    (Y,X_1)=(Y,X_2)=(Y,X_3)=(Y,I_\infty)=(Y,Z)=(Z,X_3)=(Z,X_4)=(Z,I_\infty)=0.
\end{equation}
This has two pairs of solutions $Y^{29}_\pm$ and $Z^{29,\pm}_\pm$. Similar discussion apply to $J_{27}$, $J_{28}$ and $J_{30}$ by symmetry, and we generate solutions $\{Z^{27}_{\pm},Y^{27,\pm}_\pm\}$, $\{Z^{28}_{\pm},Y^{28,\pm}_\pm\}$ and $\{Y^{30}_{\pm},Z^{30,\pm}_\pm\}$ respectively. We see that $\{Z^{27}_\pm,Z^{28}_\pm,Y^{29}_\pm,Y^{30}_\pm\}$ are just solutions of Schubert problems for one-loop triangles, according to the on-shell conditions. Some of the solutions coincide with $\{Y^3_\pm,Z^2_\pm\}$ in the previous example since they solve the same on-shell conditions. However, we use distinguished notations to clarify that letters in this row can be purely recovered by their Schubert solutions.

For $J_{31}$, we also need to perform IBP. The upshot is that the following relation holds 
\begin{align}
    J_{31}=\frac{r_1J_{1,1,0,1,1,0,2,0,0}}{\epsilon}&= -r_1(m_1^2 J_{1,1,0,1,1,1,1,0,1}-J_{1,1,0,1,1,0,1,0,1}-J_{1,1,0,1,1,1,1,0,0})\nonumber\\
    &+\text{lower sector integrals}
\end{align}
whose maximal cut corresponds to the following conditions
\[(Y,X_1)=(Y,X_2)=(Y,I_\infty)=(Y,Z)=(Z,X_1)=(Z,X_2)=(Z,X_3)=(Z,X_4)=0\]
generating solutions $\{Y^{31}_\pm,Z^{31,\pm}_\pm\}$ respectively.

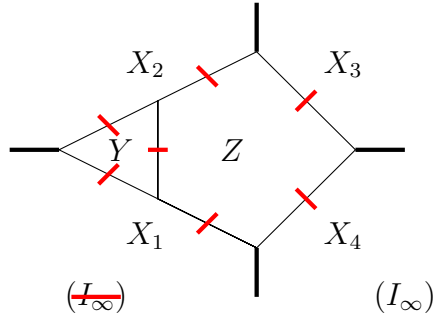
\begin{figure}[htbp]
    \centering
 \begin{tikzpicture}[scale=0.65]
 	\draw[black] (1,-2)--(3,0)--(1,2)--(-3,0)--(1,-2)--(-1,-1)--(-1,1)--(-1,-1)--cycle;
 	\draw[black,ultra thick] (-3,0)--(-4,0);
 	\draw[black,ultra thick] (3,0)--(4,0);
    \draw[black,ultra thick] (1,2)--(1,3);
    \draw[black,ultra thick] (1,-2)--(1,-3);
    \node (X1) at (0.5,0) {$Z$};
    \node (X1) at (-1.75,0) {$Y$};
    \node (X1) at (3+1,-2-1) {$(I_{\infty})$};
    \node (X1) at (-1-0.25-1,-3) {$(I_{\infty})$};
    \draw[red,ultra thick] (0.2-1,0)--(-0.2-1,0);
    \draw[red,ultra thick] (-2+0.2,0.5-0.2)--(-2-0.2,0.5+0.2);
    \draw[red,ultra thick] (-2+0.2,-0.5+0.2)--(-2-0.2,-0.5-0.2);
    \draw[red,ultra thick] (0+0.2,1.5-0.2)--(0-0.2,1.5+0.2);
    \draw[red,ultra thick] (0+0.2,-1.5+0.2)--(0-0.2,-1.5-0.2);
    \draw[red,ultra thick] (2+0.2,1+0.2)--(2-0.2,1-0.2);
    \draw[red,ultra thick] (+2-0.2,-1+0.2)--(2+0.2,-1-0.2);
    \draw[red,ultra thick] (-2.25+0.5,-3)--(-2.25-0.5,-3);
    \node (X1) at (3-0.25,-2+0.25) {$X_4$};
    \node (X1) at (3-0.25,2-0.25) {$X_3$};
    \node (X1) at (-1-0.25,-1.5-0.25) {$X_1$};
    \node (X1) at (-1-0.25,1.5+0.25) {$X_2$};
 \end{tikzpicture}
 \caption{Schubert problem for $J_{31}$}
\end{figure}

Now let us turn to the differential equations. Total differential of $J_{32}$ yields the expression
\begin{align}
 {\rm d}J_{32}&{=}\frac12{\rm d}\log W_{43}J_{27}{+}\frac12{\rm d}\log W_{57}J_{28}{+}\frac12{\rm d}\log W_{39}J_{29}{+}\frac12{\rm d}\log W_{58}J_{30}{+}{\rm d}\log \frac{W_{11}^5}{W_1^3W_3^3W_{18}^3}J_{32}\nonumber\\
 &{-}{\rm d}\log W_{20}J_{33}+\text{lower sector contributions}   
\end{align}
Letters appearing in this row of CDE are again easy to reproduce:
\begin{align}
    &(Y^{32}_+,Y^{32}_-)=(Z^{32}_+,Z^{32}_-)\propto\frac{W_{11}}{W_{18}}\nonumber\\
    &\frac{(Y^{32}_+,Y^{29}_-)(Y^{32}_-,Y^{29}_+)}{(Y^{32}_+,Y^{29}_+)(Y^{32}_-,Y^{29}_-)}=\frac{(Z^{32}_+,Z^{29,+}_+)(Z^{32}_-,Z^{29,-}_+)}{(Z^{32}_+,Z^{29,-}_+)(Z^{32}_-,Z^{29,+}_+)}= W_{39}^2,\ \,\frac{(Z^{32}_+,Z^{29,+}_+)(Z^{32}_-,Z^{29,+}_-)}{(Z^{32}_+,Z^{29,+}_-)(Z^{32}_-,Z^{29,+}_+)}= W_{20}^2
\end{align}
and $\{W_{43},W_{57},W_{58}\}$ are generated similarly from symmetry. It can be checked that no new letters appear in lower-sector contributions.

Total differentials of $\{J_{27}\cdots,J_{31}\}$ enjoy similar properties. For instance, ${\rm d}J_{29}$ reads
\begin{align}
    {\rm d}J_{29}&={\rm d}\log W_{35}J_{27}{-}{\rm d}\log W_{36} J_{28}{-}\frac12{\rm d}\log\frac{W_1W_2^2W_3^2W_5^2}{W_{18}^2} J_{29}{+}\frac12{\rm d}\log W_{38} J_{30}{-}\frac12{\rm d}\log W_{46} J_{31}\nonumber\\
    &-3{\rm d}\log W_{39} J_{32}+\frac12{\rm d}\log (W_{21}^3W_{26}) J_{33}+\text{lower sector contributions}
\end{align}
Among them, $\{W_{35},W_{36},W_{38},W_{46},W_{39}\}$ can be obtained just by cross-ratios from two Schubert problems according to where they appear in this row. For instance $W_{38}^2=\frac{(Y^{30}_+,Y^{29}_-)(Y^{30}_-,Y^{29}_+)}{(Y^{30}_+,Y^{29}_+)(Y^{32}_-,Y^{29}_-)}$, {\it etc.}. While for coefficient before $J_{33}$, we have
\begin{equation}
    \frac{(Y^{29}_+,Y^{28,+}_+)(Y^{29}_-,Y^{28,+}_-)}{(Y^{29}_-,Y^{28,+}_+)(Y^{29}_+,Y^{28,+}_-)}=W_{21}^2,\,  \frac{(Y^{29}_+,Y^{27,+}_+)(Y^{29}_-,Y^{27,+}_-)}{(Y^{29}_-,Y^{27,+}_+)(Y^{29}_+,Y^{27,+}_-)}=W_{21}W_{26}
\end{equation}
Letters from lower sectors can be generated in the same way.

In summary, in these examples we have seen that if two UT master integrals $J_a$ and $J_b$ both have unique maximal cut and thus can be associated with two individual Schubert problems, the coefficient between ${\rm d}J_a$ and $J_b$ (and vice versa) can be obtained from Schubert solutions from $J_a$ and $J_b$. On the other hand, even if $J_c$ is not associated to any individual Schubert problem, its coefficient in ${\rm d}J_a$ can still be generated by other Schubert solutions with those for $J_a$. Finally, we have not found any new letters appearing for ${\rm d}J_c$.  In this way we have reproduced all letters for sectors without bubble sub-diagrams in the double-box family. For sectors with bubble sub-diagrams, similar to discussions at one loop, it is only proper to define its Schubert problems in two dimensions, and similar problems arise when we consider combinations of Schubert problems in different dimensions~\footnote{Note that for ${\rm d}J_a$ without bubble sub-diagrams, it still receives non-trivial contributions ${\rm d}\log W_e J_e$ from sub-sectors with bubble sub-diagrams. However, in this example no new letters are produced from $W_e$ than letters constructed by $d=4$ Schubert solutions already. This property is not expected to be true in the most general cases.}. 

Before we conclude, let us briefly comment on more examples which have been studied or can be worked out in a similar manner. Note that in~\cite{He:2022tph} a variety of two-loop Feynman integrals in $D=4-2 \epsilon$ have been studied, including those with zero- and one-mass five particle kinematics. Although those computations were performed in twistor space, such symbol letters can be reproduced by similar computations of cross-ratios in the embedding space for $d=4$ as we have done above (see appendix~\ref{schubert} for more discussions). Similar higher-loop computations can be performed for integrals in other dimensions, {\it e.g.} we have initiated some Schubert analysis for six-particle two-loop integrals in general dimensions. In particular, the top sector involves some interesting double-pentagon integrals in $D=6-2\epsilon$ dimensions~\cite{Henn:2021cyv}. By analyzing maximal cuts in $d=6$, we find it to be given by an elliptic integral since there is one degree of freedom unfixed which appears in the square root of quartic (or cubic) polynomial, thus we expect that the double-pentagon integral generally involves elliptic MPL functions. However, such a Schubert analysis in $d=6$ simplifies a lot when we restrict the external hexagon kinematics to four dimensions. In this case, we find that the maximal cut indeed gives the square root $F_4$ as found in~\cite{Henn:2021cyv}, and at least some symbol letters involving $F_4$ can be obtained as cross-ratios of this and other maximal-cut solutions. A complete investigation of the alphabet for this sector (and possibly the entire integral family) is left for a future work~\footnote{We thank Johannes Henn, Yang Zhang and collaborators for useful discussions regarding this point.}.

\section{Conclusion and Outlook}\label{discussion}
In this note we have found an extension of the Schubert analysis originally proposed in twistor space for four-dimensional kinematics, to Feynman integrals in general dimensions with general internal masses. We have restricted ourselves to planar integrals which can be conveniently written in embedding space, and in all cases we have considered, symbol letters of the integral family can be obtained from various maximal-cut solutions for loop momenta of these integrals. More importantly, we have seen that some UT master integrals are naturally associated with Schubert problems, in which case the entries of their CDE have a natural interpretation as those cross-ratios of corresponding maximal-cut solutions. This works very nicely for the most general one-loop integrals and various two-loop integral families such as two-mass sunrise integrals in $2-2 \epsilon$ dimensions, and four-mass double-box integrals in $4-2 \epsilon$ dimensions. Therefore, our Schubert analysis produces much more detailed information than just the alphabet; once some ${\rm d} \log$ UT basis is known for a family, it seems that we can obtain detailed information about the complete CDE! We end with some concrete remarks on open problems and new directions indicated by our preliminary investigations.  

\paragraph{Special cases: $d{=}4$ and $d{=}3$}
Although we have considered general dimensions, it is natural to ask what is special about $d=4$ with massless propagators, where Schubert analysis can be performed in twistor space. In Appendix \ref{schubert} we review basics of the original Schubert analysis in momentum twistor space for $d=4$ kinematics, as well as connections with our general analysis here. Moreover, it is also natural to apply twistor-space Schubert analysis to $d=3$ kinematics, {\it e.g.} to 
amplitudes/integrals for ABJM theory in $d=3$~\cite{Caron-Huot:2012sos, He:2022lfz, He:2023rou, He:2023exb}. In this small interlude we present the Schubert analysis for ABJM $n=6$ symbology where we can still exploit momentum twistor variables by imposing the symplectic conditions \cite{Elvang:2014fja}
\begin{equation}\label{symp}
    \Omega_{IJ}Z_i^IZ_{i{+}1}^J=\Omega_{IJ}(A^IB^J)_i=0
\end{equation}
for any pairs of two adjacent $(Z_iZ_{i{+}1})$ and any loop momenta $(AB)_i$, which reduce external and loop dual points to $d=3$, and Schubert analysis can still be performed with constraints \eqref{symp}. In $d=3$, for one-loop level Schubert analysis, triangles are the most basic building blocks. For instance, on-shell conditions for $6$-point two-mass triangle $I_3(x_1,x_4,x_6)$ read 
\[(y-x_1)^2\sim\langle AB61\rangle=0,\ (y-x_4)^2\sim\langle AB34\rangle=0,\ (y-x_6)^2\sim\langle AB56\rangle=0\] and additionally we impose symplectic condition for $(AB)$. The only solution to these four conditions is $(AB)=(346)\cap(156)$, which gives an intersection $6$ on external line $(56)$. In exactly the same way as $d=4$ case, cross-ratios of such intersections on the line produce symbol letters. For instance, collecting intersections on $(56)$ from triangles $I_3(x_1,x_4,x_6)$, $I_3(x_3,x_4,x_6)$ and $I_2(x_2,x_4,x_6)$, we have the following $4$ points on a line:

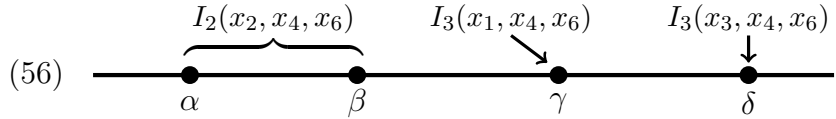
\begin{figure}[htbp]
    \centering
     \begin{tikzpicture}[scale=1.5]
 	\draw[black,ultra thick] (-3.25,0)--(3.25,0);
  \filldraw[fill=black,ultra thick] (-2.5+0.1,0) circle (0.06);
  \filldraw[fill=black,ultra thick] (-5/6-0.1,0) circle (0.06);
  \filldraw[fill=black,ultra thick] (5/6,0) circle (0.06);
  \filldraw[fill=black,ultra thick] (2.5,0) circle (0.06);
  \node (X1) at (-2.5+0.1,-0.25) {$\alpha$};
  \node (X1) at (-5/6-0.1,-0.25) {$\beta$};
  \node (X1) at (5/6,-0.25) {$\gamma$};
  \node (X1) at (2.5,-0.25) {$\delta$};
  \node (X1) at (-3.75,0) {$(56)$};
  \node[rotate=180] (X1) at (-5/3,0.25) {$\underbrace{\hspace{2.35cm}}$};
  \node (X1) at (-5/3,0.5) {\small $I_2(x_2,x_4,x_6)$};
  \node (X1) at (5/12,0.5) {\small $I_3(x_1,x_4,x_6)$};
  \node (X1) at (2.5,0.5) {\small $I_3(x_3,x_4,x_6)$};
  \draw[->,very thick] (5/12,0.35)--(5/6-0.1,0.11);
  \draw[->,very thick] (2.5,0.35)--(2.5,0.11);
 \end{tikzpicture}
    \caption{$4$ points from $d=3$ triangles; Specially, $\gamma=6$, $\delta=(56)\cap(234)$.}
    \label{fig:ABJMA1}
\end{figure}
It can be computed that the cross-ratio reads
\begin{equation}\label{minor}
\frac{[\alpha,\gamma][\beta,\delta]}{[\alpha,\delta][\beta,\gamma]}=\chi_3,\quad \text{ where}\,  [1,2]:=\langle 12I\rangle
\end{equation}
where $I$ is an arbitrary bitwistor in momentum twistor space, and we find $\chi_i=\frac{\sqrt{u_i}-\sqrt{1{-}u_i}}{\sqrt{u_i}+\sqrt{1{-}u_i}}$ (note that $\chi_1\chi_2\chi_3=1$) and $u_i=\frac{(X_i,X_{i{+}2})(X_{i{+}3},X_{i{-}1})}{(X_i,X_{i{+}3})(X_{i{+}2},X_{i{-}1})}$.

Recall that up to two loops, $n=6$ amplitudes and integrals in ABJM theory has a remarkably simple alphabet which consists of $8$ multiplicatively independent letters \cite{Caron-Huot:2012sos}:
\[\{u_1,u_2,u_3,1{-}u_1,1{-}u_2,1{-}u_3,\chi_1,\chi_2\}\]
Following the same logic as $d=4$ Schubert analysis (see appendix \ref{schubert}), we go through all possible combinations of $d=3$ triangles. The upshot is that we construct $9$ multiplicatively independent letters, $8$ of which are precisely the letters in the alphabet, while the last one is the square root $\sqrt{\frac{1-u_i}{u_i}}$ (for $i=1,2,3$ they are all proportional to each other, and also proportional to $\sqrt{x_{2,4}^2x_{4,6}^2x_{2,6}^2}=\sqrt{x_{1,3}^2x_{3,5}^2x_{1,5}^2}$ in $d=3$; they are nothing but the same leading singularity for the two three-mass triangle integrals). It would be extremely interesting to see what alphabets we get for higher point ABJM amplitudes \cite{He:2022lfz}, and consider $n=6$ and higher-point bootstrap programs similar to those for ${\cal N}=4$ SYM in $d=4$.

\begin{figure}[htbp]
\centering
\includegraphics[width=0.5\textwidth]{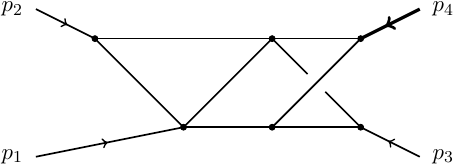}
\caption{Three-loop non-planar integral yielding the new letter}
\label{B41}
\end{figure}

\paragraph{UT integrals, Baikov ${\rm d} \log$ forms and CDE}
As we have seen, solving Schubert problems give us leading singularities associated with Feynman integrals thus it is naturally related to UT integrals and construction of $\dif\log$ forms. Since we can always study the $\dif\log$ forms and related cut conditions in other representations, it is interesting to find the connection between Schubert analysis presented in this paper with other methods like constructing $\dif\log$ forms and deriving symbol letters from intersection theory~\cite{Chen:2023kgw} or obtaining symbol alphabets from Landau singular locus~\cite{Dlapa:2023cvx}; here we will briefly comment on another possible connection: we expect that Schubert analysis must be related to a new method for constructing symbol letters directly from Baikov Gram determinants~\cite{toappear}. In appendix \ref{Baikov}, we review the direct relation between Schubert problems with $\dif\log$ forms constructed in Baikov representations, which is just the first step. The real magic seems to be that on one side, symbol letters can always be written as some cross ratios formed by solutions of Schubert problems and on the other side (in Baikov representation) they are given in terms of Baikov Gram determinants which are evaluated on the same cut conditions~\cite{toappear}! How these two seemingly very different approaches both give equivalent description of symbol letters is something worth further investigations. It would be highly desirable to find a more universal, representation-independent way for constructing symbol letters based on maximal cuts. No matter how we obtain the letters, the crucial next step is to either directly bootstrap the integrals (as MPL functions or first at symbol level) or find a way to determine the CDE these UT basis satisfy. Along this line, we have seen intriguing patterns and structures of CDE matrices from Schubert analysis once the UT basis is given. Once this can be understood better, we believe that such a ``CDE bootstrap" should be plausible, which can make both Schubert and Baikov methods more practically useful. Last but not least, a very important question is whether or not Schubert analysis can say anything about how to choose a UT basis, and we leave all these questions for future work.
\paragraph{Higher loops and non-planar integrals}
All the concrete examples in our main text is about two-loop planar integrals, but our method works for planar higher-loop integrals in the same manner provided we solve on-shell conditions from their topologies accordingly. Moreover, as briefly discussed in \cite{He:2022tph},  it is possible to consider Schubert analysis for maximal cuts of non-planar integrals, where we have to go beyond natural embedding variables for dual points. One way to proceed is that we still construct Lorentz invariants like $(\ell_1-\ell_2)^2$ or certain cross-ratios using maximal cut solutions $\{\ell_1,\ell_2\}$ of any two non-planar integrals directly in momentum space. The only subtlety is that label of two loop momenta $\ell$ in the two integrals should be compatible, which is easily guaranteed when we consider two non-planar integrals from the same integral family. For example, we have considered three-loop non-planar computation \cite{Henn:2023vbd}, where one of the master integral is as fig.\ref{B41}.
New letter $(p_4^2-s)^2-p_4^2 t$ is from its three-loop (composite) leading singularity, which is nothing but the inner product $(y_+-y_-)^2$ of its two maximal cut solutions.

\paragraph{Schubert analysis for elliptic integrals}
Finally, as we have said the original Schubert analysis has been generalized to cases when $d=4$ Feynman integrals evaluate to elliptic MPLs such as double-box integrals~\cite{Morales:2022csr} (similarly in $d=3$ for double triangles~\cite{He:2023qld}), and a natural question is how these may work in general dimensions using Schubert in embedding space. An interesting example is the following two-loop $d=2$ sunrise integral with three massive propagators (fig.\ref{fig:sunrise3}):
\begin{figure}[htbp]
    \centering
     \begin{tikzpicture}[scale=1.5]
 	\draw[black,ultra thick] (-2,0)--(-1,0);
 	\draw[black,ultra thick] (2,0)--(1,0);
 	\filldraw[color=black,fill=white,ultra thick] (0,0) circle (1);
 	\draw[black,ultra thick] (-1,0)--(1,0);
 	\node (s) at (-1.8,0.1) { $s$};
 	\node (m1) at (0,1.2) { $m_1$};
 	\node (0) at (0,0.15) {$m_3$};
 	\node (m2) at (0,-1.2) { $m_2$};
 \end{tikzpicture}
    \caption{Sunrise with three massive propagators.}
    \label{fig:sunrise3}
\end{figure}
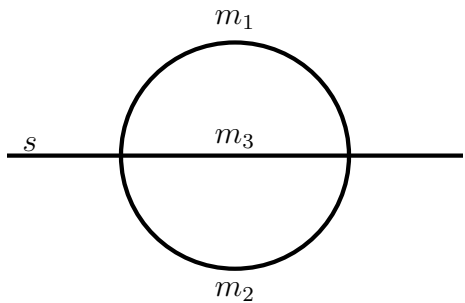
Although the mutual propagator is massive, we can still set $(y-z)^2=(Y,Z)/((Y,I_\infty)(Z,I_\infty))=m_3^2$ and recast all on-shell conditions to embedding space. Maximal residue under three on-shell conditions in embedding space gives an integral over an elliptic curve. 
It would be interesting to explore the construction of elliptic last entries of this integral from an elliptic Schubert analysis in embedding space. 




\acknowledgments

It is our pleasure to thank Yichao Tang for collaborations in the initial stage of the work, and Samuel Abreu, Ruth Britto, Johannes Henn, Zhenjie Li, Matthias Wilhelm, Xiaofeng Xu, Chi Zhang, and Yang Zhang for inspiring discussions. This work is supported in part by National Natural Science Foundation of China under Grant No. 11935013, 12047502, 12047503, 12247103, 12225510.

\appendix
\section{Schubert analysis in momentum twistor space}\label{schubert}
The original Schubert analysis was introduced in momentum twistor space, and was designed for dual conformal integrals in $\mathcal{N}=4$ SYM at the first time \cite{Yang:2022gko}. Shortly afterwards it was applied to general planar, massless propagators integrals in  QCD \cite{He:2022ctv,He:2022tph}. Most recently it also helps to predict the possible symbol entries for elliptic integral whose result was successfully bootstrapped \cite{Morales:2022csr}. In this appendix, we present a brief review of Schubert analysis in $d=4$ ($D=4-2\epsilon$) through momentum twistor variables, as well as how Schubert analysis in embedding space is related to it. 

\paragraph{A review of Schubert analysis in twistor space} Recall that for $n$ ordered, on-shell momenta $p_i$ in planar amplitudes/integrals, it is convenient to introduce $n$ {\it momentum twistors }\cite{Hodges:2009hk} $\mathbf{Z}:=Z_i^A$, with $A=1\cdots4$, following the definition:
\[Z_i=(\lambda_i^\alpha,x_i^{\alpha{\dot\alpha}}\lambda_{i\alpha})\]
where dual coordinates $x_i$ are defined by $p_i=x_{i{+}1}-x_i$ and $\lambda_i$ are spinor helicity variables associated to $p_i^2=0$. Momentum twistors trivialize both the on-shell conditions $p_i^2=0$ and the momentum conservation, and the squared distance of two dual points reads
$(x_i{-}x_j)^2=\frac{\langle i{-}1ij{-}1j\rangle}{\langle i{-}1i\rangle\langle j{-}1j\rangle}$. Here Pl\"ucker $\langle ijkl\rangle$ is the basic $SL(4)$ invariant $\langle ijkl\rangle:=\epsilon_{ABCD}Z_i^AZ_j^BZ_k^CZ_l^D$. Each dual point $x_i$ (and also embedding vector $X_i^I$) is mapped to a line $(i{-}1i)$ in momentum twistor space, and dual loop momentum $y$ ($Y^I$ in embedding space) is related to a bitwistor $(AB)$ as well. Consequently, propagator $(y-x_i)^2$ is rewritten as $\frac{\langle ABi{-}1i\rangle}{\langle AB\rangle\langle i{-}1i\rangle}$. This is our basic playground for $d=4$ Schubert analysis.

The most initial and important example for Schubert analysis, as first spotted in \cite{Hodges:2010kq,conference}, is the famous four-mass box integral and its symbol. 
\begin{align}\label{integrand1}
    \mathcal{S}\left(\begin{tikzpicture}[baseline={([yshift=-.5ex]current bounding box.center)},scale=0.15]
                \draw[black,thick] (0,5)--(-5,5)--(-5,0)--(0,0)--cycle;
                \draw[black,thick] (1.93,5.52)--(0,5)--(0.52,6.93);
                \draw[black,thick] (1.93,-0.52)--(0,0)--(0.52,-1.93);
                \draw[black,thick] (-6.93,5.52)--(-5,5)--(-5.52,6.93);
                \draw[black,thick] (-6.93,-0.52)--(-5,0)--(-5.52,-1.93);
                \filldraw[black] (1.93,6) node[anchor=west] {{$j{-}1$}};
                \filldraw[black] (0.52,6.93) node[anchor=south] {{$i$}};
                \filldraw[black] (1.93,-1) node[anchor=west] {{$j$}};
                \filldraw[black] (0.52,-1.93) node[anchor=north] {{$k{-}1$}};
                \filldraw[black] (-6.93,6) node[anchor=east] {{$l$}};
                \filldraw[black] (-5.52,6.93) node[anchor=south] {{$i{-}1$}};
                \filldraw[black] (-6.93,-1) node[anchor=east] {{$l{-}1$}};
                \filldraw[black] (-5.52,-1.93) node[anchor=north] {{$k$}};
            \end{tikzpicture}\right)= \frac1{2\Delta_{i,j,k,l}} \biggl(v\otimes \frac{z_{i,j,k,l}}{\bar z_{i,j,k,l}}+u\otimes \frac{1-\bar z_{i,j,k,l}}{1-z_{i,j,k,l}} \biggr)
\end{align}
with the definition 
\begin{equation}\label{delta}
u{=}\frac{\langle i{-}1ij{-}1j\rangle \langle k{-}1kl{-}1l\rangle}{\langle i{-}1i k{-}1k\rangle\langle  j{-}1j l{-}1l\rangle},\ v{=}\frac{\langle i{-}1il{-}1l\rangle \langle j{-}1jk{-}1k\rangle}{\langle i{-}1i k{-}1k\rangle\langle  j{-}1j l{-}1l\rangle},\Delta_{i,j,k,l}=\sqrt{(1{-}u{-}v)^2-4u v}
\end{equation} 
and $z_{i,j,k,l} \bar z_{i,j,k,l}=u,\ (1{-}z_{i,j,k,l})(1{-}\bar z_{i,j,k,l})=v$. All the four multiplicatively independent symbol letters of the integral can be obtained by the following configuration
\begin{center}
\begin{tikzpicture}[scale=0.6]
\draw[black,ultra thick](-4,2)--(-4,-2);
\draw[black,ultra thick](-2,2)--(-2,-2);
\draw[black,ultra thick](0,2)--(0,-2);
\draw[black,ultra thick](2,2)--(2,-2);
\draw[blue,thick](-4.5,1)--(2.5,1);
\draw[blue,thick](-4.5,-1)--(2.5,-1);
\filldraw[blue] (2.5,1) node[anchor=west] {{$(AB)_1$}};
\filldraw[blue] (2.5,-1) node[anchor=west] {{$(AB)_2$}};
\filldraw[black] (-4,2) node[anchor=south] {{$i{-}1$}};
\filldraw[black] (-4,-2) node[anchor=north] {{$i$}};
\filldraw[black] (-2,2) node[anchor=south] {{$j{-}1$}};
\filldraw[black] (-2,-2) node[anchor=north] {{$j$}};
\filldraw[black] (0,2) node[anchor=south] {{$k{-}1$}};
\filldraw[black] (0,-2) node[anchor=north] {{$k$}};
\filldraw[black] (2,2) node[anchor=south] {{$l{-}1$}};
\filldraw[black] (2,-2) node[anchor=north] {{$l$}};
\filldraw[blue]  (-4,1) circle [radius=2pt];
\filldraw[blue]  (-4,-1) circle [radius=2pt];
\filldraw[blue]  (-2,1) circle [radius=2pt];
\filldraw[blue]  (-2,-1) circle [radius=2pt];
\filldraw[blue]  (0,1) circle [radius=2pt];
\filldraw[blue]  (0,-1) circle [radius=2pt];
\filldraw[blue]  (2,1) circle [radius=2pt];
\filldraw[blue]  (2,-1) circle [radius=2pt];
\filldraw[blue] (-4,1) node[anchor=north east] {{$\alpha_1$}};
\filldraw[blue] (-4,-1) node[anchor=north east] {{$\alpha_2$}};
\filldraw[blue] (-2,1) node[anchor=north east] {{$\beta_1$}};
\filldraw[blue] (-2,-1) node[anchor=north east] {{$\beta_2$}};
\filldraw[blue] (0,1) node[anchor=north east] {{$\gamma_1$}};
\filldraw[blue] (0,-1) node[anchor=north east] {{$\gamma_2$}};
\filldraw[blue] (2,1) node[anchor=north east] {{$\delta_1$}};
\filldraw[blue] (2,-1) node[anchor=north east] {{$\delta_2$}};
\end{tikzpicture}
\end{center}
Here $(AB)_i$ are two solutions of {\it Schubert problem} 
\[\langle ABi{-}1i\rangle=\langle  ABj{-}1j\rangle=\langle  ABk{-}1k\rangle=\langle  ABl{-}1l\rangle=0\]
associated to four-mass box. Cross-ratios from the intersections $\{\alpha_i,\beta_i,\gamma_i,\delta_i\}_{i=1,2}$ yield the four letters by
\begin{align}
\frac{[\alpha_1,\beta_1][\gamma_1,\delta_1]}{[\alpha_1,\gamma_1][\beta_1,\delta_1]}=z_{i,j,k,l},\frac{[\alpha_1,\delta_1][\gamma_1,\beta_1]}{[\alpha_1,\gamma_1][\beta_1,\delta_1]}=1{-}z_{i,j,k,l}\nonumber\\
\frac{[\alpha_2,\beta_2][\gamma_2,\delta_2]}{[\alpha_2,\gamma_2][\beta_2,\delta_2]}=\bar z_{i,j,k,l},\frac{[\alpha_2,\delta_2][\gamma_2,\beta_2]}{[\alpha_2,\gamma_2][\beta_2,\delta_2]}=1{-}\bar{z}_{i,j,k,l}
\end{align}
where $[Z_1,Z_2]$ for momentum twistors $Z_1$ and $Z_2$ is defined as \eqref{minor}.

\begin{figure}[htbp]
	\centering
	\includegraphics{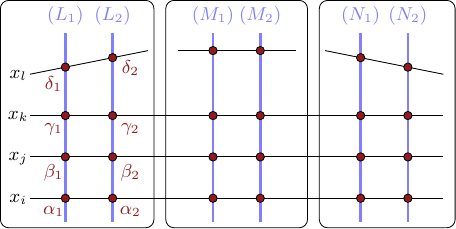}
	\caption{Combining Schubert problems for triple of four-mass boxes when analyzing $\mathcal{S}(\includegraphics{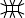})$. They share three external points in common.}
	\label{fig: Schubert intersection 3 boxes}
	\end{figure}

Move to more complicated integrals containing different sub-topologies, each of which correspond to individual Schubert problems, we should combine distinct Schubert problems and construct symbol letters, once those different Schubert problems share same external lines/points. As an illustration, let us recall the generalization of third entries for double-box integral, for which we constructed``$6$-point configurations" on external lines by combining triples of four-mass boxes as fig.\ref{fig: Schubert intersection 3 boxes}. 
Each horizontal line stands for an external dual point $x_{i}$ for a four-mass box, and each pair of vertical lines in the three boxes represents the solutions of one-loop Schubert problem for the box. The upshot is that besides cross-ratios from the blue lines,  on $x_i$ $x_j$ and $x_k$ we can construct $9$ independent cross-ratios, consisting of three rational letters proportional to $\mathcal{G}^{a,b,0}_{a,b,0}$ for $\{a,b\}\in\{\{i,j\},\{j,k\},\{i,k\}\}$, as well as six odd letters as
\begin{align}
    &\left\{\frac{\mathcal{G}^{a,0}_{b,0}+\sqrt{-\mathcal{G}_{a,b,0}^{a,b,0}\mathcal{G}^0_0}}{\mathcal{G}^{a,0}_{b,0}-\sqrt{-\mathcal{G}_{a,b,0}^{a,b,0}\mathcal{G}^0_0}}\right\},\, \{a,b\}\in\{\{i,j\},\{j,k\},\{i,k\}\}\label{hexlasts}\\
&\left\{\frac{\mathcal{G}^{a,b,0}_{a,c,0}+\sqrt{\mathcal{G}_{a,b,0}^{a,b,0}\mathcal{G}_{a,c,0}^{a,c,0}}}{\mathcal{G}^{a,b,0}_{a,c,0}-\sqrt{\mathcal{G}_{a,b,0}^{a,b,0}\mathcal{G}_{a,c,0}^{a,c,0}}}\right\},\,  \{a,b,c\}\in\{\{i,j,k\},\{j,k,i\},\{k,i,j\}\}
\end{align}
where $\mathcal{G}^0_0$ reads the Gram determinant of fully massive hexagon, and they are all reasonable candidates for the third entries of $12$-point double-box integral. In the main text we also talked about $6$-point configuration from solutions $Y^{i,j}$ in embedding space \eqref{6pt}, and we picked letters \eqref{type2} for our discussion, which is \eqref{hexlasts} in this special $d=4$ case. After we reveal the relation between Schubert analysis in embedding space and twistor space, we can see that the configuration \eqref{6pt} is a 
natural generalization of $6$-intersection configuration in fig.\ref{fig: Schubert intersection 3 boxes}.

\paragraph{Schubert analysis in twistor space and in embedding space}
Let us move to the relation between the Schubert analysis in embedding space for $d=4$ and that in twistor space. In both cases, to perform Schubert analysis, we generate maximal-cut solutions and construct invariants from intersections/solutions to predict possible symbol letters. Therefore, the only task is to figure out how the geometrical invariants from intersection points in momentum twistor space are related to invariants from $Y_\pm$ in $4+2$-dimensional embedding space. 

To do this, suppose we are combining two  Schubert problems, whose four solutions $\{(AB)_i\}_{i=1,\cdots4}$ intersect with two external lines $x_i$ and $x_j$ simultaneously. Moreover, we suppose each solution $(AB)_i$ corresponds to a vector $Y^I_i$ in $(4+2)$-dimensional embedding space.
\begin{center}
\begin{tikzpicture}[scale=0.8]
\draw[blue,thick](-4,2)--(-4,-2);
\draw[blue,thick](-2,2)--(-2,-2);
\draw[blue,thick](0,2)--(0,-2);
\draw[blue,thick](2,2)--(2,-2);
\draw[black,ultra thick](-4.5,1)--(2.5,1);
\draw[black,ultra thick](-4.5,-1)--(2.5,-1);
\filldraw[black] (2.5,1) node[anchor=west] {{$x_i$}};
\filldraw[black] (2.5,-1) node[anchor=west] {{$x_j$}};
\filldraw[blue] (-4,2) node[anchor=south] {{$(AB)_1$}};
\filldraw[blue] (-2,2) node[anchor=south] {{$(AB)_2$}};
\filldraw[blue] (0,2) node[anchor=south] {{$(AB)_3$}};
\filldraw[blue] (2,2) node[anchor=south] {{$(AB)_4$}};
\filldraw[blue]  (-4,1) circle [radius=2pt];
\filldraw[blue]  (-4,-1) circle [radius=2pt];
\filldraw[blue]  (-2,1) circle [radius=2pt];
\filldraw[blue]  (-2,-1) circle [radius=2pt];
\filldraw[blue]  (0,1) circle [radius=2pt];
\filldraw[blue]  (0,-1) circle [radius=2pt];
\filldraw[blue]  (2,1) circle [radius=2pt];
\filldraw[blue]  (2,-1) circle [radius=2pt];
\filldraw[blue] (-4,1) node[anchor=north east] {{$\alpha_1$}};
\filldraw[blue] (-4,-1) node[anchor=north east] {{$\alpha_2$}};
\filldraw[blue] (-2,1) node[anchor=north east] {{$\beta_1$}};
\filldraw[blue] (-2,-1) node[anchor=north east] {{$\beta_2$}};
\filldraw[blue] (0,1) node[anchor=north east] {{$\gamma_1$}};
\filldraw[blue] (0,-1) node[anchor=north east] {{$\gamma_2$}};
\filldraw[blue] (2,1) node[anchor=north east] {{$\delta_1$}};
\filldraw[blue] (2,-1) node[anchor=north east] {{$\delta_2$}};
\end{tikzpicture}
\end{center}
Therefore from the configuration we have four invariants $\{\frac{[\alpha_i,\delta_i][\gamma_i,\beta_i]}{[\alpha_i,\gamma_i][\beta_i,\delta_i]},\frac{[\alpha_i,\beta_i][\gamma_i,\delta_i]}{[\alpha_i,\gamma_i][\beta_i,\delta_i]}\}_{i=1,2}$. Parametrizing $\gamma_i=A_i\alpha_i+B_i\beta_i$, $\delta_i=C_i\alpha_i+D_i\beta_i$ due to the colinearity of intersections, four cross-ratios then read $\{\frac{A_iD_i}{B_iC_i},1{-}\frac{A_iD_i}{B_iC_i}\}_{i=1,2}$. 

Now let us consider another two invariants:
\[\left\{\frac{(1 \cdot 4)(2 \cdot 3)}{(1\cdot3)(2\cdot4)}, \frac{(1\cdot2)(3\cdot4)}{(1\cdot3)(2\cdot4)}\right\}=\left\{\frac{(Y_1,Y_4)(Y_2,Y_3)}{(Y_1,Y_3)(Y_2,Y_4)},\frac{(Y_1,Y_2)(Y_3,Y_4)}{(Y_1,Y_3)(Y_2,Y_4)}\right\}\]
with $(i\cdot j):=\langle A_i,B_i,A_j,B_j\rangle$ being the Pl\"uckers formed by two lines/bitwistors $(AB)_i$ and $(AB)_j$. We see that due to the correspondence of $(AB)_i$ and $Y_i^I$, we do nothing but rewrite cross-ratios of $Y_i$ into momentum twistor space. On the other hand, cross-ratios of lines are related to original invariants by
\begin{equation}\label{rel}\frac{(1\cdot4)(2\cdot3)}{(1\cdot3)(2\cdot4)}=\frac{\langle\alpha_1,
\alpha_2,\delta_1,\delta_2\rangle\langle\beta_1,\beta_2,\gamma_1,\gamma_2\rangle}{\langle\alpha_1,
\alpha_2,\gamma_1,\gamma_2\rangle\langle\beta_1,\beta_2,\delta_1,\delta_2\rangle}=\frac{A_1A_2D_1D_2}{B_1B_2C_1C_2}=\prod_{i=1}^2\frac{[\alpha_i,\delta_i][\gamma_i,\beta_i]}{[\alpha_i,\gamma_i][\beta_i,\delta_i]}\end{equation}
and similar for the other one. Therefore, cross-ratios formed by the embedding solutions turn out to be products of two cross-ratios formed by the intersections. This is the most naive relation between two kinds of invariants we construct.

However, as we have seen in twistor space, letters from all possible cross-ratios by intersections are always of great redundancy, and we need to get rid of the unnecessary ones. In \cite{Morales:2022csr} a nice property was spotted; to cover the third entries of the integrals, we only need combinations of box Schubert problems that share {\it three} external lines at the same time. It can be straightforward proved that three $6$-point configurations on $x_i$, $x_j$ or $x_k$ in fig.\ref{fig: Schubert intersection 3 boxes} are in fact related by $GL(2)$ transformations. Therefore, letters constructed from the configurations are actually independent of which external line they are from; on each shared external line, we get {\it identical} cross-ratios!  
Generalizing this property, for arbitrary $k$ Schubert problems, we will consider their combinations, if and only if on all their shared external lines the $2k$-point configurations are identical and give the same cross-ratios. We have gone through multiple examples up to three loops in $d=4$ and found that this rule always works in removing spurious ones and giving the correct alphabet.

Going back to the expression \eqref{rel} and adopting this rule, we obtain
\[\left\{\frac{(Y_1,Y_4)(Y_2,Y_3)}{(Y_1,Y_3)(Y_2,Y_4)},\frac{(Y_1,Y_2)(Y_3,Y_4)}{(Y_1,Y_3)(Y_2,Y_4)}\right\}=\left\{\biggl(\frac{[\alpha_i,\delta_i][\gamma_i,\beta_i]}{[\alpha_i,\gamma_i][\beta_i,\delta_i]}\biggr)^2,\biggl(\frac{[\alpha_i,\beta_i][\gamma_i,\delta_i]}{[\alpha_i,\gamma_i][\beta_i,\delta_i]}\biggr)^2\right\}\]
for either $i=1$ or $2$, {\it i.e.} cross-ratios formed by the embedding solutions turn out to be square of cross-ratios formed by the intersections. Therefore, we can view the analysis in embedding space as a natural generalization for the original twistor-space Schubert analysis. 

Finally we remark that although for Schubert analysis in momentum twistor space the selection rule works very well, it remains unclear in general which Schubert problems can be combined and give non-trivial letters in embedding space. This is an important open problem since it is closely related to the following question: if we have UT basis elements $I_a, I_b$ associated with two Schubert problems, 
when do we expect {\it zero coefficient} of $I_b$ in the CDE of $d I_a$? A probable answer is that it happens if and only if the two Schubert problems from $I_a$ and $I_b$ can not be combined, thus a general selection rule of this form will be very important.

\section{Baikov ${\rm d}\log$ forms and Schubert problems}\label{Baikov}

There is evidence that Schubert problems presented in the main text are related to $\mathrm{d}\log$ form integrals which can be seen clearly in Baikov representation. More explicitly, we expect that one Schubert problem can correspond to one $\mathrm{d}\log$ form the simple poles of which will serve as the solution of this Schubert problem. Let us review the Baikov representation and describe how to construct $\mathrm{d}\log$ forms as well as how they are related to Schubert problems.

\paragraph{Review of Baikov representations} The main idea of Baikov representation is to transform the familiar Feynman integrals with respect to loop momentum $l^{\mu}$ to the integration of independent scalar products $l\cdot q_{i}$ where $q_{i}$ is either loop momentum or external momentum. Since denominators of propagators are linear combinations of scalar products, the integral can be further transformed into integration of a complete set of such denominators which we will denote as $\{x_{i}\}$. Then the general structure of standard Baikov representation will be
\begin{equation}
    I=C\int\frac{\dif x_1\cdots\dif x_N}{x_1^{a_1}\cdots x_N^{a_N}} \left[ P(x_1,\cdots,x_N) \right]^{(D-L-E-1)/2}
\end{equation}
where $C$ is some prefactor depending on external kinematics; $L$ is the number of loops, $E$ is the number of independent external momenta and $N=L(L+1)/2+LE$ is the number of independent scalar products $l\cdot q_{i}$; $D=d-2\epsilon$ contains the dimensional regularization parameter and $d$ is some integer as in the main text. 

Here $P(x_1,\cdots,x_N)$ is called Baikov polynomial which is a polynomial of $x_i$'s. It is actually a determinant of Gram matrix of loop momenta and external momenta. When $a_{i}>0$, then $x_{i}$ is in the denominator and it corresponds to the propagators of a Feynman diagram, while $a_{i}<0$, it is an ISP in the numerator. Some of the ISPs can be integrated out directly in the representation of higher loop integral and the new representation is related to the so-called loop-by-loop Baikov representations:
\begin{equation}
    I=C\int\frac{\dif x_1\cdots\dif x_m}{x_1^{a_1}\cdots x_m^{a_m}} \prod_{i}\left[ P_{i}(x_1,\cdots,x_m) \right]^{\gamma_{i}}
\end{equation}
where $P_{i}$ are all polynomials and they are actually minors of original determinant $P$. $\gamma_{i}$ are powers related to dimension $d$ and $m\le N$.  

One advantage of Baikov representation is that maximal cut conditions become straightforward. When $a_i=1$, the maximal cut condition for cutting this propagator is exactly taking the residue of $x_i$ at $x_i=0$, and more cut conditions can be put on once we construct a $\mathrm{d}\log$ form 
\begin{equation}
    \int\prod\left[ P_{i}(x_1,\cdots,x_m) \right]^{\epsilon}\dif\log f_{1}\wedge \ldots \dif\log f_{m}
\end{equation}
by requiring $f_{i}=0 \text{ or }\infty$. Note that $\epsilon$ is an infinitesimal parameter. When $\epsilon=0$, $\prod\left[P\right]^{\epsilon}$ goes to 1 and we go back to the integer dimension. A set of such conditions $\{f_{i}=0 \text{ or } \infty\}, i=1,\ldots,m$ then gives us a Schubert problem. Now we discuss in more details following the track of the main text. 

\paragraph{One-loop integrals} Let's begin with one-loop case. For an $n$-gon integral $I$, we can define its Baikov representation and the corresponding Baikov polynomial $P(\mathbf{x})$ where $\mathbf{x}=(x_1,\ldots
,x_n,x_{0}\equiv 1)$ and $x_{i}=D_{i}$ is the denominator of propagator\footnote{We abuse the notation a little here by using $x_i$ instead of $D_{i}$ to denote the denominator of propagator for the simplicity of expression. But it should be clear in this appendix, $x_i$ has a different meaning with that in main text.}. $x_0$ actually corresponds to $(Y,I_{\infty})$ in the embedding space. Due to the simplicity of one-loop integrals, $P(\mathbf{x})$ is a quadratic polynomial of $\mathbf{x}$, so it can be written as $P(\mathbf{x})=\mathbf{x}\cdot Q \cdot \mathbf{x}$ where $Q$ is a matrix. Actually, all symbol letters can be written as expressions of minors of this single matrix. We further denote $Q_{M,N}$ as the minor of $Q$ by taking rows in $M$ and columns in $N$. For example, $Q_{12,13}$ is a minor of $Q$ by taking 1-st, 2-nd rows and 1-st, 3-rd columns. However, $Q_{0,0}$ will correspond to taking the last-row and last-column element of $Q$.

The construction of $\dif\log$ is rather straightforward in one-loop case, since there is only one nontrivial polynomial $P(\mathbf{x})$ in the representation and its power $p$ is $(D-L-E-1)/2=(d-n-1)/2-\epsilon$ for an $n$-gon integral. For arbitrary positive integer $n$, we can always pick up some integer dimension so that $p=-\epsilon$ or $p=-1/2-\epsilon$. In the first case, the $\dif\log$ form reads
\begin{equation}\label{eq:oneloopdlog1}
    \int\left[P(\mathbf{x})\right]^{-\epsilon}\dif\log x_1 \wedge \dif\log x_2 \wedge \ldots \dif\log x_n \, .
\end{equation}
In the second case, the $\dif\log$ form can be constructed as\footnote{When constructing a $\dif\log$ form for one variable, all other integration variables are taken as constants. Once one variable is constructed, the remaining expression must be free of this variable. Following one construction order, finally the wedge product will guarantee that this is actually a $\dif\log$ form for all the integration variables.}
\begin{equation}\label{eq:oneloopdlog2}
    \int\left[P(\mathbf{x})\right]^{-\epsilon}\frac{\sqrt{P(\mathbf{x})|_{x_1=0}}\dif x_1}{x_1\sqrt{P(\mathbf{x})}} \wedge \frac{\sqrt{P(\mathbf{x})|_{x_{1,2}=0}}\dif x_2}{x_2\sqrt{P(\mathbf{x})|_{x_1=0}}} \wedge \ldots \frac{\sqrt{P(x_n)|_{x_n=0}}\dif x_n}{x_n\sqrt{P(x_n)}} 
\end{equation}
where $P(x_n)$ is $P(\mathbf{x})$ after setting $x_1,\ldots,x_{n-1}$ to 0. In above construction, we have exploited that the following form is in general a $\dif\log$ form:
\begin{equation}
    \frac{\sqrt{(c-c_0)(c-c_1)}}{(z-c)\sqrt{(z-c_0)(z-c_1)}}\dif z=-\dif\log\frac{1+\sqrt{\frac{(c_1-c)(c_0-z)}{(c_0-c)(c_1-z)}}}{1-\sqrt{\frac{(c_1-c)(c_0-z)}{(c_0-c)(c_1-z)}}}
\end{equation}
There is another kind of general formula which is also useful in later construction
\begin{equation}
    \frac{\dif z}{\sqrt{(z-c_0)(z-c_1)}}=\dif\log\frac{1+\sqrt{\frac{c_0-z}{c_1-z}}}{1-\sqrt{\frac{c_0-z}{c_1-z}}} \, .
\end{equation}
In both \eqref{eq:oneloopdlog1} and \eqref{eq:oneloopdlog2}, the $\dif\log$ form corresponds to one type of Schubert problem, that is
\begin{equation}
    x_i=0, i=1,\ldots,n
\end{equation}
This directly corresponds to the first line in \eqref{eq:cutcondition}. The second line in \eqref{eq:cutcondition} actually has a similar origin. We can introduce an ISP by hand for one-loop family and this can always be done in Baikov representation. If we integrate this ISP out we will arrive at the original representation. However, we can also keep it and construct a $\dif
\log$ form for this ISP and it turns out that finally the condition will be the second line of \eqref{eq:cutcondition}. This phenomenon will be more common in higher loop case as we will see below.

Now for one-loop integral family, we can drive the general form of its canonical differential equation system and get the the symbol alphabet \cite{Jiang:2023qnl}. It is summarized as the followings. When $n$ is even,
\begin{equation}\label{DE1}
   \begin{aligned}
       \mathrm{d} I=\epsilon &\left( -\mathrm{d}\log\frac{Q_{0,0}}{Q_{\emptyset,\emptyset}} I + \frac{1}{2}\sum_{i\neq 0}\mathrm{d}\log\frac{Q_{i,0}+\sqrt{Q_{i,i}Q_{0,0}}}{Q_{i,0}-\sqrt{Q_{i,i}Q_{0,0}}} I^{i} \right. \\
       &\left. + \frac{1}{4}\sum_{0<i<j}\mathrm{d}\log\frac{Q_{i0,j0}+\sqrt{-Q_{0,0}Q_{ij0,ij0}}}{Q_{i0,j0}-\sqrt{-Q_{0,0}Q_{ij0,ij0}}} I^{i,j}\right) \, ,
   \end{aligned}
\end{equation}
and when $n$ is odd,
\begin{equation}\label{DE2}
   \begin{aligned}
       \mathrm{d} I=\epsilon & \left( -\mathrm{d}\log\frac{Q_{0,0}}{Q_{\emptyset,\emptyset}} I + \frac{1}{2}\sum_{i\neq 0}\mathrm{d}\log\frac{Q_{i,0}+\sqrt{-Q_{\emptyset,\emptyset}Q_{i0,i0}}}{Q_{i,0}-\sqrt{-Q_{\emptyset,\emptyset}Q_{i0,i0}}} I^{i} \right. \\
       &\left. + \frac{1}{4}\sum_{0<i<j}\mathrm{d}\log\frac{Q_{i,j}+\sqrt{-Q_{\emptyset,\emptyset}Q_{ij,ij}}}{Q_{i,j}-\sqrt{-Q_{\emptyset,\emptyset}Q_{ij,ij}}} I^{i,j}\right) \, .
   \end{aligned}
\end{equation}
We define $Q_{\emptyset,\emptyset}\equiv \mathcal{G}(p_{1},p_{2},\ldots p_{n-1})=-(-\frac{1}{2})^{n-1}\mathcal{G}(X_1,\ldots,X_{n},I_{\infty})$. 
All these minors of $Q$ are actually Gram determinants of momenta. They are related to $\mathcal{G}^A_B$ in the main text by the following formula
\begin{equation}
    \begin{aligned}
        Q_{0,0}&=(-\frac{1}{2})^{n}\mathcal{G}_{0}^{0}, \, \, Q_{i0,i0}=-(-\frac{1}{2})^{n+1}\mathcal{G}_{i,0}^{i,0}, \, \, Q_{ij0,ij0}=(-\frac{1}{2})^{n+2}\mathcal{G}_{i,j,0}^{i,j,0}, \, \\
        Q_{i,i}&=(-\frac{1}{2})^{n}\mathcal{G}_{i}^{i}, \,\, Q_{ij,ij}=-(-\frac{1}{2})^{n+1}\mathcal{G}_{i,j}^{i,j}, \, \,
        Q_{ijk,ijk}=(-\frac{1}{2})^{n+2}\mathcal{G}_{i,j,k}^{i,j,k}.
    \end{aligned}
\end{equation}
For the convenience of notation, the minors of $Q$ have been normalized so that they will not contain $Q_{\emptyset,\emptyset}$\footnote{Actually they will depend on powers of $Q_{\emptyset,\emptyset}$ according to the definition. But it is convenient to set this factor to 1 in calculation and rescale it back to get the right expression.}. The seemingly complicated coefficients like $(-1/2)^{n}$ all originate from the mismatch between the definition of product $(X,Y)$ in \eqref{eq:dotproduct} and Gram determinant of momenta in \eqref{eq:grammatrix}. At last, the off-diagonal minors like $Q_{i,0}$ are all related to diagonal ones by
\begin{equation}\label{defQab}
    Q_{Ma,Mb}^{2}=Q_{Ma,Ma}Q_{Mb,Mb}-Q_{M,M}Q_{Mab,Mab}
\end{equation}
This identity has many names in the literature, for example Sylvester identity in~\cite{Dlapa:2021qsl}, Jacobi identity in~\cite{Dlapa:2023cvx} or the special case of Lewis-Carroll identity in~\cite{fomin_introduction_2021}. Therefore we still have $Q_{a,b}=(-\frac12)^n\mathcal{G}^a_b$,{\it etc.}\footnote{Note that since \eqref{defQab} only fix $Q_{Ma,Mb}$ up to a sign, here we can alternatively choose $Q_{a,b}=-(-\frac12)^n\mathcal{G}^a_b$ ,which results in an extra sign before ${\rm d}\log$, so the sign before 1/2 and 1/4 in \eqref{DE1} and \eqref{DE2} can also be minus in an explicit calculation. Nevertheless this sign can be easily determined numerically.}. All the $(-\frac12)$ factors are then exactly canceled in the expression \eqref{DE1} and \eqref{DE2}, and the expressions show perfect agreement with \eqref{DES1}, \eqref{DES2} from Schubert analysis once we set $\{x_i,y_i,z_i\}=\{-1,\frac12,\frac14\}$ in these two formula. Note that this result also agrees with diagrammatic coactions \cite{Abreu:2017mtm}, where the authors obtained these symbol letters from cut integrals as well.

\paragraph{Higher-loop Schubert problems from Baikov representations} Next we move to higher-loop cases, where ISP will always appear. We can just construct $\dif\log$ form as the one-loop case and determine its corresponding Schubert problems. We first analyze the two-loop sunrise \eqref{sunrise} in the main text. We can actually construct the following $\dif\log$ forms in the loop-by-loop Baikov representation\footnote{We can actually construct another $\dif\log$ form $F_{2}^{\prime}=\dif\log D_{1}\wedge \dif\log D_{2}\wedge\dif\log D_{3}\wedge\frac{\sqrt{\lambda(m_1^2,m_2^2,s)}\dif D_{5}}{D_{5}\sqrt{\lambda(D_{5}+m_2^2,m_1^2,s)}}$ in a different loop-by-loop representation. However, it will correspond to the same Schubert problem as $F_{2}$.}:
\begin{equation}
    \begin{aligned}
        F_{1}&=\dif\log \frac{\dif D_{2}}{\sqrt{\lambda(D_1,D_2,D_4)-4m_1^2 D_{2}}}\wedge\dif\log D_{1}\wedge \dif\log D_{3}\wedge \frac{\dif D_{4}}{\sqrt{\lambda(D_4+m_1^2,m_2^2,s)}}, \\
        F_{2}&=\dif\log D_{1}\wedge \dif\log D_{2}\wedge\dif\log D_{3}\wedge\frac{\sqrt{\lambda(m_1^2,m_2^2,s)}\dif D_{4}}{D_{4}\sqrt{\lambda(D_{4}+m_1^2,m_2^2,s)}}, \\
        F_{3}&=\dif\log D_{1}\wedge \dif\log D_{2}\wedge\dif\log D_{3}\wedge\frac{\dif D_{4}}{\sqrt{\lambda(D_{4}+m_1^2,m_2^2,s)}}, \\
        F_{4}&=\dif\log D_{1}\wedge \dif\log D_{2}\wedge\dif\log D_{3}\wedge\frac{\dif D_{5}}{\sqrt{\lambda(D_{5}+m_2^2,m_1^2,s)}}\, .
    \end{aligned}
\end{equation}
Note that $F_2$, $F_3$ and $F_4$ belong to top sector and $F_1$ corresponds to the only non-trivial sub-sector. We can see directly that $F_{2}$ gives the following conditions:
\begin{equation}
   F_2:\quad D_1=D_2=D_3=D_4=0
\end{equation}
which is exactly the Schubert problem \eqref{prob1}. $F_3$ and $F_4$ decides the following conditions respectively
\begin{equation}
    \begin{aligned}
        F_3:\quad D_1=D_2=D_3=0,\,D_4=\infty , \\
        F_4:\quad D_1=D_2=D_3=0, \, D_5=\infty .
    \end{aligned}
\end{equation}
Since in projective space, sending $D_4=(Z,X_1)/(Z,I_\infty)$ (or $D_5=(Y,X_2)/(Y,I_\infty)$) to infinity is achieved by setting $(Z,I_{\infty})$ (or $(Y,I_{\infty})$) to 0. Thus these two conditions actually correspond to the Schubert problems \eqref{prob2} and \eqref{prob3}. At last, $F_1$ gives the conditions
\begin{equation}
    F_1:\quad D_1=D_3=0,\, D_2=D_4=\infty \, .
\end{equation}
It corresponds to the Schubert problem \eqref{prob4} for the same reason.

Next we move to the double-box and double-triangle integrals \eqref{doublebox}. In both cases we can work in the maxima-cut Baikov representations for the corresponding sector since by performing maximal cut we are considering the real non-trivial part belonging to this sector. Maximal cut eliminate the sub-sector integrals. And the corresponding Schubert problems will always cut all the propagators for the same reason. So for double box,
\begin{equation}
    D_1=D_2=D_3=D_4=D_5=D_6=D_7=0
\end{equation}
which corresponds to
\begin{equation}
    (Y,X_1)=(Y,X_2)=(Y,X_3)=(Y,Z)=(Z,X_3)=(Z,X_4)=(Z,X_1)=0
\end{equation}
will always be part of the whole condition. What we need to do is constructing $\dif\log$ forms for the remaining ISPs and give additional conditions. We can actually construct 3 such $\dif\log$ forms directly\footnote{Actually, in a different loop-by-loop Baikov representation we can construct another $\dif\log$ form $E_1^{\prime}=\frac{r_1\dif D_9}{D_9\sqrt{Q(D_9)}}$ which is equivalent to $E_1$. It will also correspond to the same Schubert problem, like the discussion in the main text.}
\begin{equation}
    \begin{aligned}
        E_1=\frac{r_1\dif D_8}{D_8\sqrt{P(D_8)}}  , \,
        E_2=\frac{r_2\dif D_8}{\sqrt{P(D_8)}} , \,
        E_3=\frac{r_4\dif D_9}{\sqrt{Q(D_9)}} \, .
    \end{aligned}
\end{equation}
$P(D_8)$ and $Q(D_9)$ are quadratic polynomials of $D_8$ and $D_9$. They are actually two Gram determinants $G(k_1,p_1,p_2,p_3)$ and $G(k_2,p_1,p_2,p_3)$ after setting $D_1=D_2=\ldots=D_7=0$ where $k_1$ and $k_2$ are the corresponding two loop momenta of the double box. $E_1$ decides the following condition
\begin{equation}
    E_1: \quad D_8 =0
\end{equation}
which corresponds to $J_{1}$ in the Schubert problem \eqref{eq:dbprob}. $E_2$ and $E_3$ decides
\begin{equation}
    E_2:\quad D_8 =\infty, \quad E_3:\quad D_9 =\infty
\end{equation}
which corresponds to $J_3$ and $J_2$ respectively in \eqref{eq:dbprob}. As for $J_4$, the equivalent $\dif\log$ form is different from above three $\dif\log$ forms. In two different loop-by-loop Baikov representations we can construct different $\dif\log$ forms which both correspond to $J_4$. However, they correspond to different Schubert problems. So $J_4$ is not associated with a definite Schubert problem. And this agrees with the analysis in main text.

For double triangle, the common cut condition will be
\begin{equation}
    D_1=D_2=D_4=D_5=D_7=0
\end{equation}
which corresponds to
\begin{equation}
    (Y,X_1)=(Y,X_2)=(Y,Z)=(Z,X_3)=(Z,X_4)=0 \, .
\end{equation}
Now we only consider the $\dif\log$ construction for ISPs. For double triangle, the $\dif\log$ construction will involve a non-linear variable transformation as pointed out in \cite{Dlapa:2021qsl}. This makes it difficult to present them in a simple way so we don't show the expression explicitly. Nevertheless, we find that each of $J_{27}\cdots J_{32}$ yields an individual ${\rm d}\log$, and corresponds to Schubert problem in the same way as before. While for $J_{33}$, its Baikov representation reads different ${\rm d}\log$-forms from different approaches again, therefore does not correspond to an individual Schubert problem just like $J_4$.

\section{Symbol letters for four-mass double-box family}\label{4mletters}
When discussing the double-box integral family in section 3, we adopt the following notation for square roots and symbol letters according to \cite{He:2022ctv}. For the square roots
\begin{equation}\label{defi:r1-5}
\begin{aligned}
r_1^2=& s^2 t^2-2 s t m_1^2 m_3^2+m_1^4 m_3^4-2 s t m_2^2 m_4^2-2 m_1^2 m_2^2 m_3^2 m_4^2+m_2^4 m_4^4,\\
r_2^2=& s^2-2 s m_1^2+m_1^4-2 s m_2^2-2 m_1^2 m_2^2+m_2^4,\\
r_3^2=& t^2-2 t m_1^2+m_1^4-2 t m_4^2-2 m_1^2 m_4^2+m_4^4,\\
r_4^2=& s^2-2 s m_3^2+m_3^4-2 s m_4^2-2 m_3^2 m_4^2+m_4^4,\\
r_5^2=& t^2-2 t m_2^2+m_2^4-2 t m_3^2-2 m_2^2 m_3^2+m_3^4.\\
r_6^2=& s^2+2 s t+t^2-2 s m_1^2-2 t m_1^2+m_1^4-2 s m_3^2-2 t m_3^2+2 m_1^2 m_3^2+m_3^4-4 m_2^2 m_4^2,\\
r_7^2=& s^2+2 s t+t^2-2 s m_2^2-2 t m_2^2+m_2^4-4 m_1^2 m_3^2-2 s m_4^2-2 t m_4^2+2 m_2^2 m_4^2+m_4^4,\\
r_8^2=& s^2 t^2-2 s^2 t m_1^2+s^2 m_1^4+2 s t m_1^2 m_4^2-2 s m_1^4 m_4^2-2 s t m_2^2 m_4^2+2 s m_1^2 m_2^2 m_4^2\nonumber\\&-4 s m_1^2 m_3^2 m_4^2+m_1^4 m_4^4-2 m_1^2 m_2^2 m_4^4+m_2^4 m_4^4,\\
r_9^2=& s^2 t^2-2 s^2 t m_2^2+s^2 m_2^4-2 s t m_1^2 m_3^2+2 s t m_2^2 m_3^2+2 s m_1^2 m_2^2 m_3^2-2 s m_2^4 m_3^2\nonumber\\&+m_1^4 m_3^4-2 m_1^2 m_2^2 m_3^4+m_2^4 m_3^4-4 s m_2^2 m_3^2 m_4^2,\\
r_{10}^2=& s^2 t^2-2 s^2 t m_3^2+2 s t m_2^2 m_3^2-4 s m_1^2 m_2^2 m_3^2+s^2 m_3^4-2 s m_2^2 m_3^4+m_2^4 m_3^4\nonumber\\&-2 s t m_2^2 m_4^2+2 s m_2^2 m_3^2 m_4^2-2 m_2^4 m_3^2 m_4^2+m_2^4 m_4^4,\\
r_{11}^2=& s^2 t^2-2 s t m_1^2 m_3^2+m_1^4 m_3^4-2 s^2 t m_4^2+2 s t m_1^2 m_4^2-4 s m_1^2 m_2^2 m_4^2+2 s m_1^2 m_3^2 m_4^2\nonumber\\&-2 m_1^4 m_3^2 m_4^2+s^2 m_4^4-2 s m_1^2 m_4^4+m_1^4 m_4^4.
\end{aligned}
\end{equation}
As for the letters, we have even ones and odd ones. The even letters are
\begin{equation}
\begin{aligned}
W_1=& m_1^2,\,
W_2= m_2^2,\,
W_3= m_3^2,\,
W_4= m_4^2,\,
W_5= s,\,
W_6= t,\\
W_7=& r_5^2,\,
W_8= r_3^2,\,
W_9= r_2^2,\,
W_{10}= r_4^2,\,
W_{11}= r_7^2,\,
W_{12}= r_6^2,\\
W_{13}=& r_1^2,\,
W_{14}= r_8^2,\,
W_{15}= r_9^2,\,
W_{16}= r_{10}^2,\,
W_{17}= r_{11}^2,\\
W_{18}=& s^2 t+s t^2-s t m_1^2-s t m_2^2+s m_1^2 m_2^2-s t m_3^2-s m_1^2 m_3^2-t m_1^2 m_3^2+m_1^4 m_3^2\\&+t m_2^2 m_3^2-m_1^2 m_2^2 m_3^2+m_1^2 m_3^4-s t m_4^2+t m_1^2 m_4^2-s m_2^2 m_4^2-t m_2^2 m_4^2\\&-m_1^2 m_2^2 m_4^2+m_2^4 m_4^2+s m_3^2 m_4^2-m_1^2 m_3^2 m_4^2-m_2^2 m_3^2 m_4^2+m_2^2 m_4^4,
\end{aligned}
\end{equation}
The odd letters that contain one square root are,
\begin{equation}
\begin{aligned}
W_{19}=& \frac{f_{19}+r_6}{f_{19}-r_6},\,
W_{20}= \frac{f_{20}+r_7}{f_{20}-r_7},\,
W_{21}= \frac{f_{21}+r_2}{f_{21}-r_2},\,
W_{22}= \frac{f_{22}+r_4}{f_{22}-r_4},\\
W_{23}=& \frac{f_{23}+r_5}{f_{23}-r_5},\,
W_{24}= \frac{f_{24}+r_3}{f_{24}-r_3},\,
W_{25}= \frac{f_{25}+r_1}{f_{25}-r_1},\\
W_{26}=& \frac{f_{26}+\left(m_1^2-m_2^2\right) r_2}{f_{26}-\left(m_1^2-m_2^2\right) r_2},\,
W_{27}= \frac{f_{27}+\left(m_3^2-m_4^2\right) r_4}{f_{27}-\left(m_3^2-m_4^2\right) r_4},\\
W_{28}=& \frac{f_{28}+\left(m_2^2-m_3^2\right) r_5}{f_{28}-\left(m_2^2-m_3^2\right) r_5},\,
W_{29}= \frac{f_{29}+\left(m_1^2-m_4^2\right) r_3}{f_{29}-\left(m_1^2-m_4^2\right) r_3},\\
W_{30}=& \frac{f_{30}+\left(m_1^2 m_3^2-m_2^2 m_4^2\right) r_1}{f_{30}-\left(m_1^2 m_3^2-m_2^2 m_4^2\right) r_1},\\
W_{31}=& \frac{f_{31}+\left(s t-s m_3^2+m_2^2 m_3^2-m_2^2 m_4^2\right) r_{10}}{f_{31}-\left(s t-s m_3^2+m_2^2 m_3^2-m_2^2 m_4^2\right) r_{10}},\\
W_{32}=& \frac{f_{32}+\left(-s t+m_1^2 m_3^2+s m_4^2-m_1^2 m_4^2\right) r_{11}}{f_{32}-\left(-s t+m_1^2 m_3^2+s m_4^2-m_1^2 m_4^2\right) r_{11}},\\
W_{33}=& \frac{f_{33}+\left(s t-s m_1^2+m_1^2 m_4^2-m_2^2 m_4^2\right) r_8}{f_{33}-\left(s t-s m_1^2+m_1^2 m_4^2-m_2^2 m_4^2\right) r_8},\\
W_{34}=& \frac{f_{34}+\left(-s t+s m_2^2+m_1^2 m_3^2-m_2^2 m_3^2\right) r_9}{f_{34}-\left(-s t+s m_2^2+m_1^2 m_3^2-m_2^2 m_3^2\right) r_9},
\end{aligned}
\end{equation}
where
\begin{equation}
\begin{aligned}
f_{19}=& -s-t+m_1^2+m_3^2,\,
f_{20}= -s-t+m_2^2+m_4^2,\,
f_{21}= s-m_1^2-m_2^2,\\
f_{22}=& s-m_3^2-m_4^2,\,
f_{23}= t-m_2^2-m_3^2,\,
f_{24}= t-m_1^2-m_4^2,\\
f_{25}=& s t-m_1^2 m_3^2-m_2^2 m_4^2,\\
f_{26}=& -s m_1^2+m_1^4-s m_2^2-2 m_1^2 m_2^2+m_2^4,\\
f_{27}=& -s m_3^2+m_3^4-s m_4^2-2 m_3^2 m_4^2+m_4^4,\\
f_{28}=& -t m_2^2+m_2^4-t m_3^2-2 m_2^2 m_3^2+m_3^4,\\
f_{29}=& -t m_1^2+m_1^4-t m_4^2-2 m_1^2 m_4^2+m_4^4,\\
f_{30}=& -s t m_1^2 m_3^2+m_1^4 m_3^4-s t m_2^2 m_4^2-2 m_1^2 m_2^2 m_3^2 m_4^2+m_2^4 m_4^4,\\
f_{31}=& s^2 t^2-2 s^2 t m_3^2+2 s t m_2^2 m_3^2-2 s m_1^2 m_2^2 m_3^2+s^2 m_3^4-2 s m_2^2 m_3^4+m_2^4 m_3^4\\&-2 s t m_2^2 m_4^2+2 s m_2^2 m_3^2 m_4^2-2 m_2^4 m_3^2 m_4^2+m_2^4 m_4^4,\\
f_{32}=& s^2 t^2-2 s t m_1^2 m_3^2+m_1^4 m_3^4-2 s^2 t m_4^2+2 s t m_1^2 m_4^2-2 s m_1^2 m_2^2 m_4^2\\&+2 s m_1^2 m_3^2 m_4^2-2 m_1^4 m_3^2 m_4^2+s^2 m_4^4-2 s m_1^2 m_4^4+m_1^4 m_4^4,\\
f_{33}=& s^2 t^2-2 s^2 t m_1^2+s^2 m_1^4+2 s t m_1^2 m_4^2-2 s m_1^4 m_4^2-2 s t m_2^2 m_4^2+2 s m_1^2 m_2^2 m_4^2\\&-2 s m_1^2 m_3^2 m_4^2+m_1^4 m_4^4-2 m_1^2 m_2^2 m_4^4+m_2^4 m_4^4,\\
f_{34}=& s^2 t^2-2 s^2 t m_2^2+s^2 m_2^4-2 s t m_1^2 m_3^2+2 s t m_2^2 m_3^2+2 s m_1^2 m_2^2 m_3^2-2 s m_2^4 m_3^2\\&+m_1^4 m_3^4-2 m_1^2 m_2^2 m_3^4+m_2^4 m_3^4-2 s m_2^2 m_3^2 m_4^2.
\end{aligned}
\end{equation}

The odd letters that contain 2 square roots are
\begin{align*}
W_{35}&= \frac{f_{35}+r_2 r_4}{f_{35}-r_2 r_4},\,
W_{36}= \frac{f_{36}+r_2 r_5}{f_{36}-r_2 r_5},\,
W_{37}= \frac{f_{37}+r_2 r_6}{f_{37}-r_2 r_6},\,
W_{38}= \frac{f_{38}+r_2 r_3}{f_{38}-r_2 r_3},\\
W_{39}&= \frac{f_{39}+r_2 r_7}{f_{39}-r_2 r_7},\,
W_{40}= \frac{f_{40}+r_3 r_4}{f_{40}-r_3 r_4},\,
W_{41}= \frac{f_{41}+r_4 r_6}{f_{41}-r_4 r_6},\,
W_{42}= \frac{f_{42}+r_4 r_5}{f_{42}-r_4 r_5},\\
W_{43}&= \frac{f_{43}+r_4 r_7}{f_{43}-r_4 r_7},\,
W_{44}= \frac{f_{44}+r_2 r_9}{f_{44}-r_2 r_9},\,
W_{45}= \frac{f_{45}+r_2 r_8}{f_{45}-r_2 r_8},\,
W_{46}= \frac{f_{46}+r_1 r_2}{f_{46}-r_1 r_2},\\
W_{47}&= \frac{f_{47}+r_2 r_{10}}{f_{47}-r_2 r_{10}},\,
W_{48}= \frac{f_{48}+r_2 r_{11}}{f_{48}-r_2 r_{11}},\,
W_{49}= \frac{f_{49}+r_4 r_{11}}{f_{49}-r_4 r_{11}},\,
W_{50}= \frac{f_{50}+r_4 r_{10}}{f_{50}-r_4 r_{10}},\\
W_{51}&= \frac{f_{51}+r_1 r_4}{f_{51}-r_1 r_4},\,
W_{52}= \frac{f_{52}+r_4 r_8}{f_{52}-r_4 r_8},\,
W_{53}= \frac{f_{53}+r_4 r_9}{f_{53}-r_4 r_9},\,
W_{54}= \frac{f_{54}+r_3 r_5}{f_{54}-r_3 r_5},\\
W_{55}&= \frac{f_{55}+r_5 r_6}{f_{55}-r_5 r_6},\,
W_{56}= \frac{f_{56}+r_3 r_6}{f_{56}-r_3 r_6},\,
W_{57}= \frac{f_{57}+r_5 r_7}{f_{57}-r_5 r_7},\,
W_{58}= \frac{f_{58}+r_3 r_7}{f_{58}-r_3 r_7},\\
W_{59}&= \frac{f_{59}+r_1 r_5}{f_{59}-r_1 r_5},\,
W_{60}= \frac{f_{60}+r_5 r_9}{f_{60}-r_5 r_9},\,
W_{61}= \frac{f_{61}+r_5 r_{10}}{f_{61}-r_5 r_{10}},\,
W_{62}= \frac{f_{62}+r_1 r_3}{f_{62}-r_1 r_3},\\
W_{63}&= \frac{f_{63}+r_3 r_8}{f_{63}-r_3 r_8},\,
W_{64}= \frac{f_{64}+r_3 r_{11}}{f_{64}-r_3 r_{11}},\,
W_{65}= \frac{f_{65}+r_1 r_8}{f_{65}-r_1 r_8},\,
W_{66}= \frac{f_{66}+r_1 r_9}{f_{66}-r_1 r_9},\\
W_{67}&= \frac{f_{67}+r_1 r_{10}}{f_{67}-r_1 r_{10}},\,
W_{68}= \frac{f_{68}+r_1 r_{11}}{f_{68}-r_1 r_{11}}.
\end{align*}
where

\begin{align*}
f_{35}=& -s^2-2 s t+s m_1^2+s m_2^2+s m_3^2+m_1^2 m_3^2-m_2^2 m_3^2+s m_4^2-m_1^2 m_4^2+m_2^2 m_4^2,\\
f_{36}=& -s t+t m_1^2-s m_2^2-t m_2^2-m_1^2 m_2^2+m_2^4+s m_3^2-m_1^2 m_3^2-m_2^2 m_3^2+2 m_2^2 m_4^2,\\
f_{37}=& -s^2-s t+2 s m_1^2+t m_1^2-m_1^4+s m_2^2-t m_2^2+m_1^2 m_2^2+s m_3^2-m_1^2 m_3^2\nonumber\\&-m_2^2 m_3^2+2 m_2^2 m_4^2,\\
f_{38}=& -s t-s m_1^2-t m_1^2+m_1^4+t m_2^2-m_1^2 m_2^2+2 m_1^2 m_3^2+s m_4^2-m_1^2 m_4^2-m_2^2 m_4^2,\\
f_{39}=& -s^2-s t+s m_1^2-t m_1^2+2 s m_2^2+t m_2^2+m_1^2 m_2^2-m_2^4+2 m_1^2 m_3^2+s m_4^2\nonumber\\&-m_1^2 m_4^2-m_2^2 m_4^2,\\
f_{40}=& -s t+s m_1^2+t m_3^2-m_1^2 m_3^2-s m_4^2-t m_4^2-m_1^2 m_4^2+2 m_2^2 m_4^2-m_3^2 m_4^2+m_4^4,\\
f_{41}=& -s^2-s t+s m_1^2+2 s m_3^2+t m_3^2-m_1^2 m_3^2-m_3^4+s m_4^2-t m_4^2-m_1^2 m_4^2\nonumber\\&+2 m_2^2 m_4^2+m_3^2 m_4^2,\\
f_{42}=& -s t+s m_2^2-s m_3^2-t m_3^2+2 m_1^2 m_3^2-m_2^2 m_3^2+m_3^4+t m_4^2-m_2^2 m_4^2-m_3^2 m_4^2,\\
f_{43}=& -s^2-s t+s m_2^2+s m_3^2-t m_3^2+2 m_1^2 m_3^2-m_2^2 m_3^2+2 s m_4^2+t m_4^2-m_2^2 m_4^2\nonumber\\&+m_3^2 m_4^2-m_4^4,\\
f_{44}=& -s^2 t+s t m_1^2-s^2 m_2^2-s t m_2^2-s m_1^2 m_2^2+s m_2^4+s m_1^2 m_3^2-m_1^4 m_3^2+s m_2^2 m_3^2\nonumber\\&+2 m_1^2 m_2^2 m_3^2-m_2^4 m_3^2+2 s m_2^2 m_4^2,\\
f_{45}=& -s^2 t-s^2 m_1^2-s t m_1^2+s m_1^4+s t m_2^2-s m_1^2 m_2^2+2 s m_1^2 m_3^2+s m_1^2 m_4^2-m_1^4 m_4^2\nonumber\\&+s m_2^2 m_4^2+2 m_1^2 m_2^2 m_4^2-m_2^4 m_4^2,\\
f_{46}=& -s^2 t+s t m_1^2+s t m_2^2-2 s m_1^2 m_2^2+s m_1^2 m_3^2-m_1^4 m_3^2+m_1^2 m_2^2 m_3^2+s m_2^2 m_4^2\nonumber\\&+m_1^2 m_2^2 m_4^2-m_2^4 m_4^2,\\
f_{47}=& -s^2 t+s t m_1^2+s t m_2^2-2 s m_1^2 m_2^2+s^2 m_3^2-s m_1^2 m_3^2-2 s m_2^2 m_3^2-m_1^2 m_2^2 m_3^2\nonumber\\&+m_2^4 m_3^2+s m_2^2 m_4^2+m_1^2 m_2^2 m_4^2-m_2^4 m_4^2,\\
f_{48}=& -s^2 t+s t m_1^2+s t m_2^2-2 s m_1^2 m_2^2+s m_1^2 m_3^2-m_1^4 m_3^2+m_1^2 m_2^2 m_3^2+s^2 m_4^2\nonumber\\&-2 s m_1^2 m_4^2+m_1^4 m_4^2-s m_2^2 m_4^2-m_1^2 m_2^2 m_4^2,\\
f_{49}=& -s^2 t+s t m_3^2+s m_1^2 m_3^2-m_1^2 m_3^4-s^2 m_4^2-s t m_4^2+s m_1^2 m_4^2+2 s m_2^2 m_4^2\nonumber\\&-s m_3^2 m_4^2+2 m_1^2 m_3^2 m_4^2+s m_4^4-m_1^2 m_4^4,\\
f_{50}=& -s^2 t-s^2 m_3^2-s t m_3^2+2 s m_1^2 m_3^2+s m_2^2 m_3^2+s m_3^4-m_2^2 m_3^4+s t m_4^2+s m_2^2 m_4^2\nonumber\\&-s m_3^2 m_4^2+2 m_2^2 m_3^2 m_4^2-m_2^2 m_4^4,\\
f_{51}=& -s^2 t+s t m_3^2+s m_1^2 m_3^2-m_1^2 m_3^4+s t m_4^2+s m_2^2 m_4^2-2 s m_3^2 m_4^2+m_1^2 m_3^2 m_4^2\nonumber\\&+m_2^2 m_3^2 m_4^2-m_2^2 m_4^4,\\
f_{52}=& -s^2 t+s^2 m_1^2+s t m_3^2-s m_1^2 m_3^2+s t m_4^2-2 s m_1^2 m_4^2+s m_2^2 m_4^2-2 s m_3^2 m_4^2\nonumber\\&-m_1^2 m_3^2 m_4^2+m_2^2 m_3^2 m_4^2+m_1^2 m_4^4-m_2^2 m_4^4,\\
f_{53}=& -s^2 t+s^2 m_2^2+s t m_3^2+s m_1^2 m_3^2-2 s m_2^2 m_3^2-m_1^2 m_3^4+m_2^2 m_3^4+s t m_4^2\nonumber\\&-s m_2^2 m_4^2-2 s m_3^2 m_4^2+m_1^2 m_3^2 m_4^2-m_2^2 m_3^2 m_4^2,\\
f_{54}=& -2 s t-t^2+t m_1^2+t m_2^2-m_1^2 m_2^2+t m_3^2+m_1^2 m_3^2+t m_4^2+m_2^2 m_4^2-m_3^2 m_4^2,\\
\end{align*}
\begin{align*}
    f_{55}=& -s t-t^2+t m_1^2-s m_2^2+t m_2^2-m_1^2 m_2^2+s m_3^2+2 t m_3^2-m_1^2 m_3^2+m_2^2 m_3^2\nonumber\\&-m_3^4+2 m_2^2 m_4^2,\\
f_{56}=& -s t-t^2+s m_1^2+2 t m_1^2-m_1^4+t m_3^2-m_1^2 m_3^2-s m_4^2+t m_4^2+m_1^2 m_4^2\nonumber\\&+2 m_2^2 m_4^2-m_3^2 m_4^2,\\
f_{57}=& -s t-t^2+s m_2^2+2 t m_2^2-m_2^4-s m_3^2+t m_3^2+2 m_1^2 m_3^2+m_2^2 m_3^2+t m_4^2\nonumber\\&-m_2^2 m_4^2-m_3^2 m_4^2,\\
f_{58}=& -s t-t^2-s m_1^2+t m_1^2+t m_2^2-m_1^2 m_2^2+2 m_1^2 m_3^2+s m_4^2+2 t m_4^2+m_1^2 m_4^2\nonumber\\&-m_2^2 m_4^2-m_4^4,\\
f_{59}=& -s t^2+s t m_2^2+s t m_3^2+t m_1^2 m_3^2-2 t m_2^2 m_3^2+m_1^2 m_2^2 m_3^2-m_1^2 m_3^4+t m_2^2 m_4^2\nonumber\\&-m_2^4 m_4^2+m_2^2 m_3^2 m_4^2,\\
f_{60}=& -s t^2+2 s t m_2^2-s m_2^4+s t m_3^2+t m_1^2 m_3^2+s m_2^2 m_3^2-t m_2^2 m_3^2-m_1^2 m_2^2 m_3^2\nonumber\\&+m_2^4 m_3^2-m_1^2 m_3^4-m_2^2 m_3^4+2 m_2^2 m_3^2 m_4^2,\\
f_{61}=& -s t^2+s t m_2^2+2 s t m_3^2+s m_2^2 m_3^2-t m_2^2 m_3^2+2 m_1^2 m_2^2 m_3^2-m_2^4 m_3^2-s m_3^4\nonumber\\&+m_2^2 m_3^4+t m_2^2 m_4^2-m_2^4 m_4^2-m_2^2 m_3^2 m_4^2,\\
f_{62}=& -s t^2+s t m_1^2+t m_1^2 m_3^2-m_1^4 m_3^2+s t m_4^2-2 t m_1^2 m_4^2+t m_2^2 m_4^2+m_1^2 m_2^2 m_4^2\nonumber\\&+m_1^2 m_3^2 m_4^2-m_2^2 m_4^4,\\
f_{63}=& -s t^2+2 s t m_1^2-s m_1^4+s t m_4^2+s m_1^2 m_4^2-t m_1^2 m_4^2+m_1^4 m_4^2+t m_2^2 m_4^2\nonumber\\&-m_1^2 m_2^2 m_4^2+2 m_1^2 m_3^2 m_4^2-m_1^2 m_4^4-m_2^2 m_4^4,\\
f_{64}=& -s t^2+s t m_1^2+t m_1^2 m_3^2-m_1^4 m_3^2+2 s t m_4^2+s m_1^2 m_4^2-t m_1^2 m_4^2-m_1^4 m_4^2\nonumber\\&+2 m_1^2 m_2^2 m_4^2-m_1^2 m_3^2 m_4^2-s m_4^4+m_1^2 m_4^4,\\
f_{65}=& -s^2 t^2+s^2 t m_1^2+s t m_1^2 m_3^2-s m_1^4 m_3^2-s t m_1^2 m_4^2+2 s t m_2^2 m_4^2-s m_1^2 m_2^2 m_4^2\nonumber\\&+2 s m_1^2 m_3^2 m_4^2-m_1^4 m_3^2 m_4^2+m_1^2 m_2^2 m_3^2 m_4^2+m_1^2 m_2^2 m_4^4-m_2^4 m_4^4,\\
f_{66}=& -s^2 t^2+s^2 t m_2^2+2 s t m_1^2 m_3^2-s t m_2^2 m_3^2-s m_1^2 m_2^2 m_3^2-m_1^4 m_3^4+m_1^2 m_2^2 m_3^4\nonumber\\&+s t m_2^2 m_4^2-s m_2^4 m_4^2+2 s m_2^2 m_3^2 m_4^2+m_1^2 m_2^2 m_3^2 m_4^2-m_2^4 m_3^2 m_4^2,\\
f_{67}=& -s^2 t^2+s^2 t m_3^2+s t m_1^2 m_3^2-s t m_2^2 m_3^2+2 s m_1^2 m_2^2 m_3^2-s m_1^2 m_3^4-m_1^2 m_2^2 m_3^4\nonumber\\&+2 s t m_2^2 m_4^2-s m_2^2 m_3^2 m_4^2+m_1^2 m_2^2 m_3^2 m_4^2+m_2^4 m_3^2 m_4^2-m_2^4 m_4^4,\\
f_{68}=& -s^2 t^2+2 s t m_1^2 m_3^2-m_1^4 m_3^4+s^2 t m_4^2-s t m_1^2 m_4^2+s t m_2^2 m_4^2+2 s m_1^2 m_2^2 m_4^2\nonumber\\&-s m_1^2 m_3^2 m_4^2+m_1^4 m_3^2 m_4^2+m_1^2 m_2^2 m_3^2 m_4^2-s m_2^2 m_4^4-m_1^2 m_2^2 m_4^4
\end{align*}
\bibliographystyle{utphys}
\bibliography{main}
\end{document}